\documentclass[galaxies,article,submit,moreauthors,pdflatex,10pt,a4paper]{mdpi} 
\graphicspath{{Definitions/}}
\usepackage{xcolor}
\usepackage{mathalfa}
\usepackage{hyperref}
\usepackage{amsmath,amssymb}
\usepackage{multirow}
\DeclareFontFamily{U}{rcjhbltx}{}
\DeclareFontShape{U}{rcjhbltx}{m}{n}{<->rcjhbltx}{}
\DeclareSymbolFont{hebrewletters}{U}{rcjhbltx}{m}{n}
\let\aleph\relax\let\beth\relax
\let\gimel\relax\let\daleth\relax
\DeclareMathSymbol{\aleph}{\mathord}{hebrewletters}{39}
\DeclareMathSymbol{\beth}{\mathord}{hebrewletters}{98}
\DeclareMathSymbol{\gimel}{\mathord}{hebrewletters}{103}
\DeclareMathSymbol{\daleth}{\mathord}{hebrewletters}{100}
\DeclareMathSymbol{\lamed}{\mathord}{hebrewletters}{108}
\DeclareMathSymbol{\mem}{\mathord}{hebrewletters}{109}
\DeclareMathSymbol{\ayin}{\mathord}{hebrewletters}{96}
\DeclareMathSymbol{\tsadi}{\mathord}{hebrewletters}{118}
\DeclareMathSymbol{\qof}{\mathord}{hebrewletters}{113}
\DeclareMathSymbol{\shin}{\mathord}{hebrewletters}{152}
\preto{\abstractkeywords}{\nolinenumbers}
\def\aap{A\&A\/}
\def\apjs{ApJS}
\def\apj{ApJ}
\def\aj{AJ}
\def\nat{Nat}
\def\apjl{ApJL}
\def\apss{ApSS}
\def\mnras{MNRAS}

\def\rmxaa{RevMexA\&Ap}

\def\nh{$n_{\mathrm{H}}$\/}
\def\lledd{$L/L_{\rm Edd}$}

\def\rfe{$R_{\rm FeII}$}

\def\feiiq{\rm Fe{\sc ii}$\lambda$4570\/}

\def\ltsima{$\; \buildrel < \over \sim \;$}
\def\ltsim{\lower.5ex\hbox{\ltsima}}  
\def\gtsima{$\; \buildrel > \over \sim \;$}

\def\gtsim{\lower.5ex\hbox{\gtsima}} 

\def\lya{{ Ly}$\alpha$}

\def\civ{{\sc{Civ}}\/}

\def\cm3{cm$^{-3}$\/}
\def\hb{{\sc{H}}$\beta$\/}

\def\hbbc{{\sc{H}}$\beta_{\rm BC}$\/}
\def\hbvbc{{\sc{H}}$\beta_{\rm VBC}$\/}

\def\oiiiopt{{\sc{[Oiii]}}$\lambda\lambda$4959,5007\/}

\def\siiii{Si{\sc iii]}\/}
\def\aliii{Al{\sc iii}\/}
\def\ciii{C{\sc iii]}\/}

\def\heiiuv{He{\sc{ii}}$\lambda$1640}
\def\nv{{N{\sc v}}\/}

\def\feii{{Fe\sc{ii}}\/}

\def\fe{{\sc{Fe}}\/}

\def\fe76087{{\sc [Fe vii]}$\lambda$6087\/}
\def\oiii{{\sc [Oiii]}$\lambda$5007}

\def\kms{km~s$^{-1}$}

\def\heii{{\sc H}e{\sc ii}\/}

\def\siiv{Si{\sc iv}}
\def\oiv{O{\sc iv]}$\lambda$1402\/}

\usepackage{enumerate}
\definecolor{darkorange}{rgb}{1,0.612,0}
\definecolor{aquamarine}{rgb}{0.498,1,0.8314}

\Title{Metal content of relativistically  jetted and radio-quiet quasars in the main sequence context}
\Author{{Paola Marziani}$^{1}$, Swayamtrupta Panda$^{2}$ Alice Deconto Machado$^{3}$, Ascension Del Olmo$^{3}$ }
\AuthorNames{Paola Marziani, Swayamtrupta Panda, Alice Deconto Machado, and Ascension Del Olmo}
\address{$^{1}$ \quad National Institute for Astrophysics (INAF), Astronomical Observatory of Padova, vicolo dell'Osservatorio 5, IT-35122 Padova, Italy\\
$^{2}$ \quad Laborat\'orio Nacional de Astrof\'isica, R. dos Estados Unidos, 154 - Na\c{c}\~oes, Itajub\'a - MG, 37504-364, Brazil\\
$^{3}$ \quad Instituto de Astrofis\'{\i}ca de Andaluc\'{\i}a, IAA-CSIC, {Glorieta}  de la Astronomia s/n, E-18008 Granada, Spain, chony@iaa.es (A.d.O.)   }
\corres{Correspondence: paola.marziani@inaf.it; Tel.: +39-0498293415}
\abstract{Optical and UV properties of radio-quiet  {(RQ)} and radio-loud ({RL,} relativistically "jetted") active galactic nuclei (AGN) are known to differ markedly; however, it is still unclear what is due to a sample selection and what is associated with intrinsic differences in the inner workings of their emitting regions. Chemical composition is an important { parameter related to the trends of} the quasar main sequence. Recent works suggest that in addition to physical properties such as density, column density, and ionization level, strong \feii\ emitters require very high metal content. Little is known, however, about the chemical composition of jetted radio-loud sources. In this short note, we present a pilot analysis of the chemical composition of low-$z$\ radio-loud and radio-quiet quasars.  Optical and UV spectra from ground and space were combined to allow for precise measurements of metallicity-sensitive diagnostic ratios. The comparison between radio-quiet and radio-loud was carried out for sources in the same domain of the Eigenvector 1 / main sequence parameter space. Arrays of dedicated photoionization simulations with the input of appropriate spectral energy distributions indicate that metallicity is   sub-solar for  RL AGN,  and slightly sub-solar or around solar for RQ AGN. The metal content of the broad line emitting region likely reflects a similar enrichment story for both classes of AGN not involving recent circum-nuclear or nuclear starbursts. } 
\begin{document}
%\maketitle
\section{Introduction}
%\vspace{-6pt}
%\subsection{Quasar Spectra: still open to interpretation}

\begin{figure}[t!]
\includegraphics[scale=1]{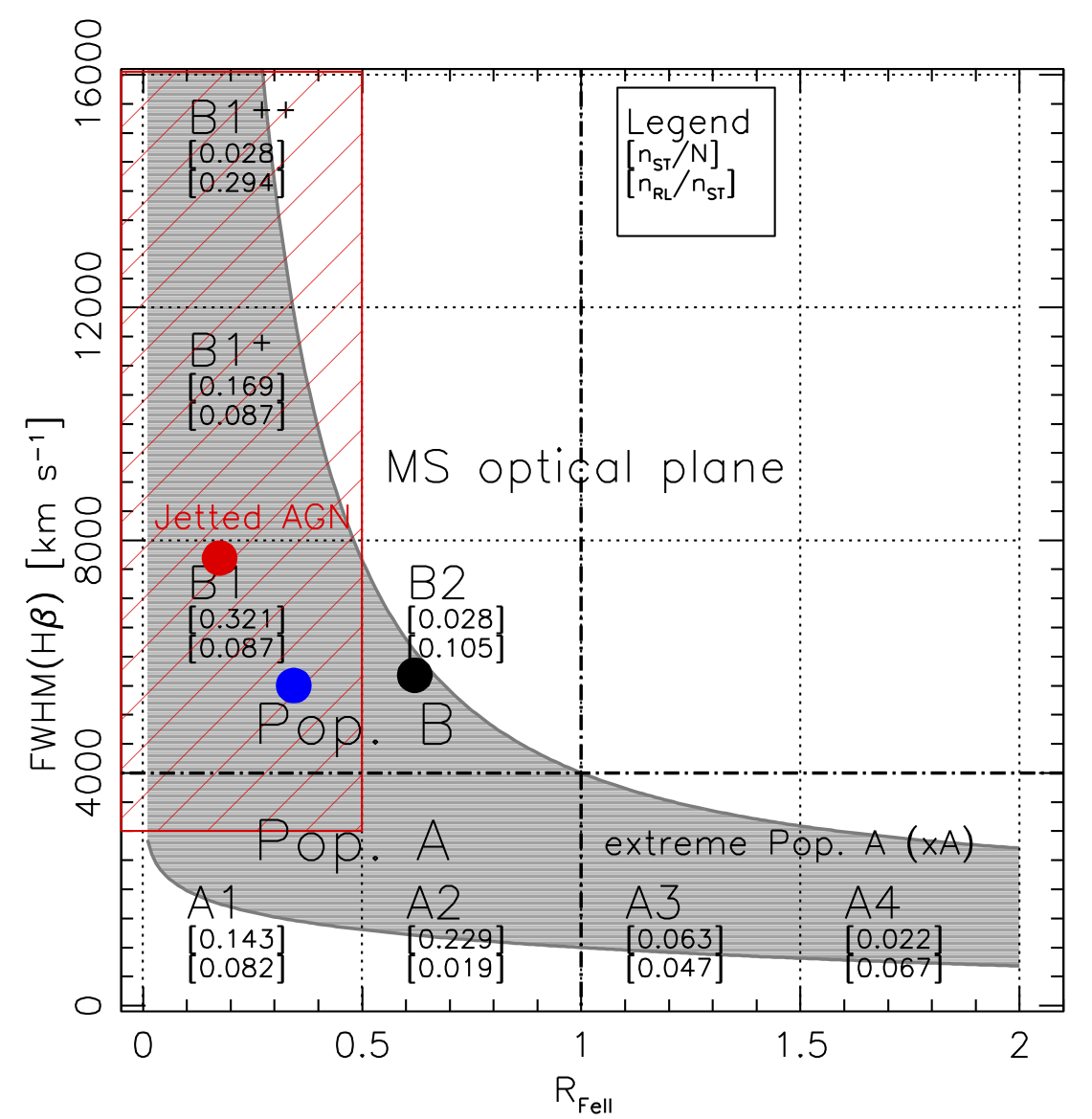}
\centering
 \caption{Sketch of the optical plane of the MS, FWHM \hb\ vs. \rfe. Numbers in square brackets provide, from top to bottom, the complete sample fractional occupation, the fraction of jetted (core-dominated + Fanaroff-Riley II) sources in each spectral bin. The shaded area identifies the region along the sequence with a high prevalence of jetted sources (c.f. Ref. \citep{marzianietal21}). The original sample is described in Ref. \cite{marzianietal03a}. Points mark the location of the composite spectra (red: RL; blue: RQ) and of PKS 0226-038 (black).}
\label{fig:ms} 
\end{figure}

Type-1 active galactic nuclei (AGN) are characterized by the presence of broad and narrow optical and UV lines (for an introduction, see e.g., \cite{netzer90,peterson97,osterbrockferland06}). AGN spectra show a mind-boggling variety of broad emission line profiles not only among different objects but also among different lines in the spectrum of the same object. For stars, the identification of optical spectral types and luminosity classes allows for the knowledge of the most relevant physical parameters, including the star's evolutionary status \citep{kaler97}. This feat is not as yet possible for AGN. Although, progress in the empirical classification has yielded a coarse contextualization of the main accretion properties, such as black hole mass, Eddington ratio, outflow prominence, spectral energy distributions, emitting region size, the metal content of the line emitting gas, etc. \citep[e.g.,][]{peterson14,duetal16a,marzianietal16,pandaetal18, pandaetal19,ferlandetal20, Panda_PhD-Thesis_2021}. The main set of correlations was derived from a principal component analysis (PCA) on a sample of several tens of quasars \citep{borosongreen92}. The importance of the Eigenvector 1 derived from the PCA has revealed itself over the years \citep{sulenticmarziani15}, leading to the definition of what has become known as the main sequence of quasars \citep{sulenticetal00a,shenho14}. The optical plane of this main sequence is identified by the line width of the HI Balmer line \hb\ (FWHM \hb) and the prominence of a singly-ionized emission, defined as the flux ratio between the FeII blends centered at $\lambda 4570$ and \hb\ itself (hereafter \rfe). The distribution of type 1, unobscured AGN takes the form of an elbow-shaped sequence in the optical plane {(see e.g., Figure \ref{fig:ms})}. The ranges of \rfe\ and FWHM \hb\ define spectral types \citep{sulenticetal02,shenho14}, and different classes  may show different occupations in the plane (Figure \ref{fig:ms}). A case in point is provided by RQ and RL \citep{zamfiretal08}: most powerful, relativistically jetted sources cluster in Population B - with moderate  \feii\ emission and relatively broad \hb\ line profiles (4000 \kms\ $\lesssim$ FWHM(\hb) $\lesssim 8000$ \kms). The broader spectral type shows higher fractions of RL; however, these spectral bins have a low prevalence, and a sizeable fraction of all type 1 AGN falls only in bin B1. This spectral type is, therefore, well suited for an inter-comparison between RQ and RL properties. In the following, we will build composite spectra for jetted and non-jetted sources of spectral type B1 (\S\ \ref{compos}). The analysis takes advantage of the main sequence (MS) correlations concerning line profiles in the spectral type B1 (\S\ \ref{analysis}). It's focused on the measurements of intensity ratios intended to be diagnostics of the metal content of the line emitting gas. Intensity ratios are interpreted using arrays of photoionization simulations covering a broad range in metallicity. Results (\S\ \ref{results}) suggest slightly sub-solar or solar metallicity and provide evidence of significant diversity in terms of chemical enrichment and evolutionary status along the main sequence (\S\ \ref{discussion}).

 \section{Composite spectra for spectral type B1 quasars}
\label{compos}

To obtain spectral data of good quality covering the full spectral ranges from 1000 \AA\ to $\approx 6000$ \AA\ (i.e., from \lya\ to \hb\ included) for the same object is still a non-trivial feat for low-$z$ quasars, as the UV coverage demands space-based observations. Here, we consider one sample of 20 RL and one sample of 16 RQ sources, all belonging to spectral type B1, in the {redshift range $\approx $ 0.002 -- 0.5  and $\approx 0.25 - 0.65 $, for RQ and RL respectively}. The absolute magnitudes derived from the quick look magnitudes in the {\em NASA Extragalactic Database} (NED), and the range is between --21 and --27 (RQ), and between --23.5  and --26.5 (RL), which correspond to bolometric luminosities in the range $\log L \sim 45 - 47$\ [erg s$^{-1}$].  The UV data are HST/FOS observations analyzed in Ref. \cite{sulenticetal07}, and the optical spectra were obtained from Ref. \cite{marzianietal03a}. The spectral similarity ensures consistency of black hole mass and Eddington ratio. We built median and average composites for the RQ and RL classes that look consistent. As the median combination is not improving the S/N, the fitting analysis was performed only on the averages. The average S/N is high with S/N$\approx$90 for the RQ composites, and $\approx 130$ and $\approx 55$ for the visual and UV ranges in the RL composite, respectively. The composite spectra are shown in Figure \ref{fig:combined}.
   
\section{Analysis}
\label{analysis}

\subsection{Line profiles}

Pop. B \hb\ profiles show a redward asymmetry modeled with a broader redshifted (FWHM $\sim$ 10000 \kms, shift at line base $\sim 1000 - 2000$ \kms) { and } a narrower Gaussian \citep{sulenticetal02,marzianietal09}.  The very broad Gaussian component is meant to represent the innermost part of the {broad line region (BLR)}, providing a simple representation of the radial stratification of the BLR in Pop. B suggested by reverberation mapping \citep{sneddengaskell07}. This component (hereafter the very broad component, VBC) has been associated with a physical region of high-ionization virialized gas and closest to the continuum source - the very broad line region, VBLR \citep{petersonferland86,brothertonetal94a,sulenticetal00c,wangli11}. While the physical properties of the VBLR line emitting gas are not well known, a decomposition of the full \hb\ profile into a symmetric, unshifted \hb\ component (\hbbc) and a \hbvbc\ provide an excellent fit to most \hb\ Pop. B profiles \citep{sulenticetal02,zamfiretal10}.

The multicomponent fits were performed using the \textsc{specfit} routine from \textsc{IRAF} \citep{kriss94}. This routine allows for the simultaneous minimum-$\chi^2$ fit of the continuum approximated by a power law and the spectral line components yielding FWHM, peak wavelength, and intensity for all line components. In the optical range, we fit the H$\beta$ profile and the \oiiiopt\ emission lines, and the  \feii\ multiplets for the composite objects.

   \subsection{Diagnostics of Metallicity and photo-ionization modeling}
\label{phys}
 
Diagnostics from the rest-frame UV spectrum take advantage of the observations of strong resonance lines that are collisionally excited \cite{negreteetal12,negreteetal13} and at least constrain density $n_\mathrm{H}$, ionization parameter $U$, and chemical abundance $Z$.  For instance,  C~IV$\lambda$1549/Ly$\alpha$, C~IV$\lambda$1549/(Si~IV + O~IV])$\lambda$1400, C~IV$\lambda$1549/He~II$\lambda$1640,  N~V$\lambda$1240/He~II$\lambda$1640 are sensitive to metallicity; and Al~III$\lambda$1860/Si~III{]}$\lambda$1892, Si~III{]}$\lambda$1892/C~III]$\lambda$1909 are sensitive to density, since inter-combination lines have a well defined critical density \citep{negreteetal12,marzianietal15}. Ratios of lines involving different ionic stages of the same element are obviously sensitive to the ionization parameter. The lines emitted from ionic species of Silicon and Aluminium deserve special attention - they are two elements greatly enhanced in Supernova ejecta \citep{chieffilimongi13}. This approach has yielded tight constraints, especially for sources radiating at high Eddington ratio \citep{negreteetal12,sniegowskaetal21,garnicaetal22}, where physical properties are consequently well-constrained because they converge toward an extreme. 

The photo-ionization code {\em Cloudy}  \cite{ferlandetal17} models the ionization, chemical, and thermal state of gas exposed to a radiation field and predicts its emission spectra and physical parameters. {\em Cloudy} simulations require inputs in terms of $n_\mathrm{H}$, $U$, $Z$, quasar spectral energy distribution (SED), and column density $N_\mathrm{c}$. The ionization parameter $U = {Q(H)}/{4 \pi r_\mathrm{BLR}^{2} c n_\mathrm{H}}$, where $Q(H) = \int_{\nu_{0}}^{\infty} {L_{\nu}}/{h\nu}$ is the number of ionizing photons, provides the ratio between photon and hydrogen number density and is dependent on the spectral energy distribution of the ionizing continuum. The simulations were carried out assuming the RL and RQ SEDs from Laor et al. \cite{laoretal97b} representative of the SED of B1 objects (work in preparation). The geometry was assumed open, plane-parallel, meaning that a slab of emitting gas is exposed to a radiation field only on one side. Arrays of {\em Cloudy} photo-ionization models\footnote{{N($n_\mathrm{H}$) $\times$ N(\textit{U}) = 425. The overall number of models includes 14 values of metallicity, and for RL and RQ SEDs, is 6,358.}} for a given metallicity $Z$\ and { column density} $N_\mathrm{c}$, constant density $n$\ and { ionization parameter} $U$\ { were}  evaluated at steps of 0.25 dex covering the ranges $7 \le \log n_\mathrm{H} \le 13$ [cm$^{-3}$], $-3 \le \log U \le 1$.  The single-value metallicity arrays were computed for $\log Z$ at { -3.0, -2.7, -2.3,} -2.0, -1.7, -1.3, -1.0, -0.7, -0.3, 0, 0.3., 0.7, 1.0, and 1.3 in solar units, i.e. from { 0.001} $Z_\odot$ to 20 $Z_\odot$.No dust and no microturbulence broadening were included in the calculations. The $Z$\ calculations are based on a single zone assumption for the BLR.  { The lowest values of the density may bias the solutions toward cases where significant \oiiiopt\ is expected; all cases with \oiii/\hb $>0.1$\ were excluded, as no broad \oiiiopt\ is observed. } A more refined analysis in the framework of the locally optimized emitting cloud model \citep{baldwinetal95,koristaetal97,Guoetal2020ApJ} is deferred to an eventual work.

\section{Results}
\label{results}

\subsection{ Composite spectra}

\begin{figure}
 \vspace{0cm}
\includegraphics[scale=0.6,angle=-90]{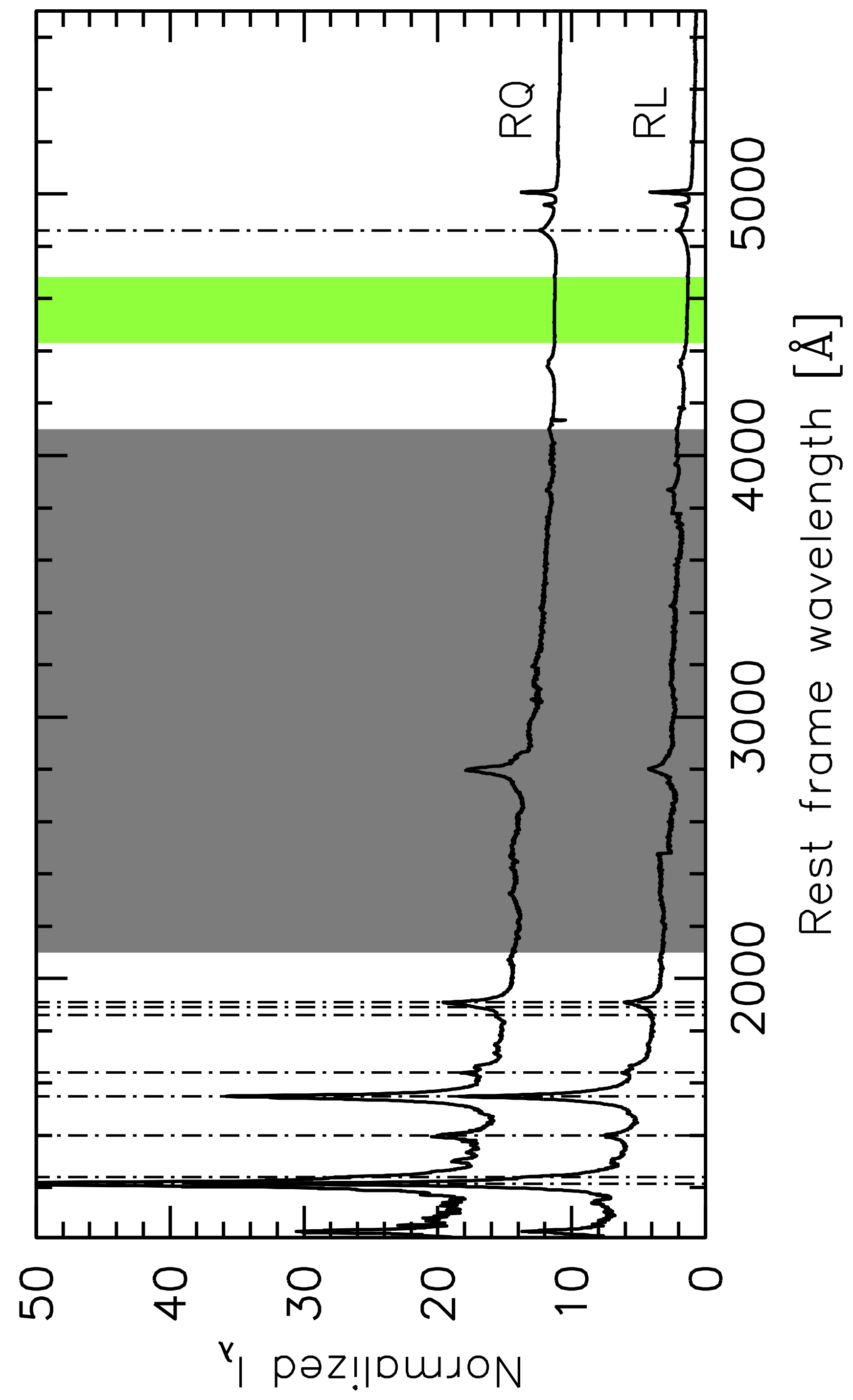} 
%\vspace{-9cm}
 \caption{The RQ and RL composite spectra covering the optical and the UV domain. The green shaded area marks the range used to measure the \feiiq\ blend. The grey-shaded region has not been considered in the analysis due to the limited number of available spectra in this region. The RL spectrum has been vertically shifted for clarity. The dot-dashed lines identify the main emission feature included in the non-linear $\chi^2$\ fitting procedures.}
\label{fig:combined} 
\end{figure}

\begin{figure}
\centering
\includegraphics[angle=-90,scale=0.55]{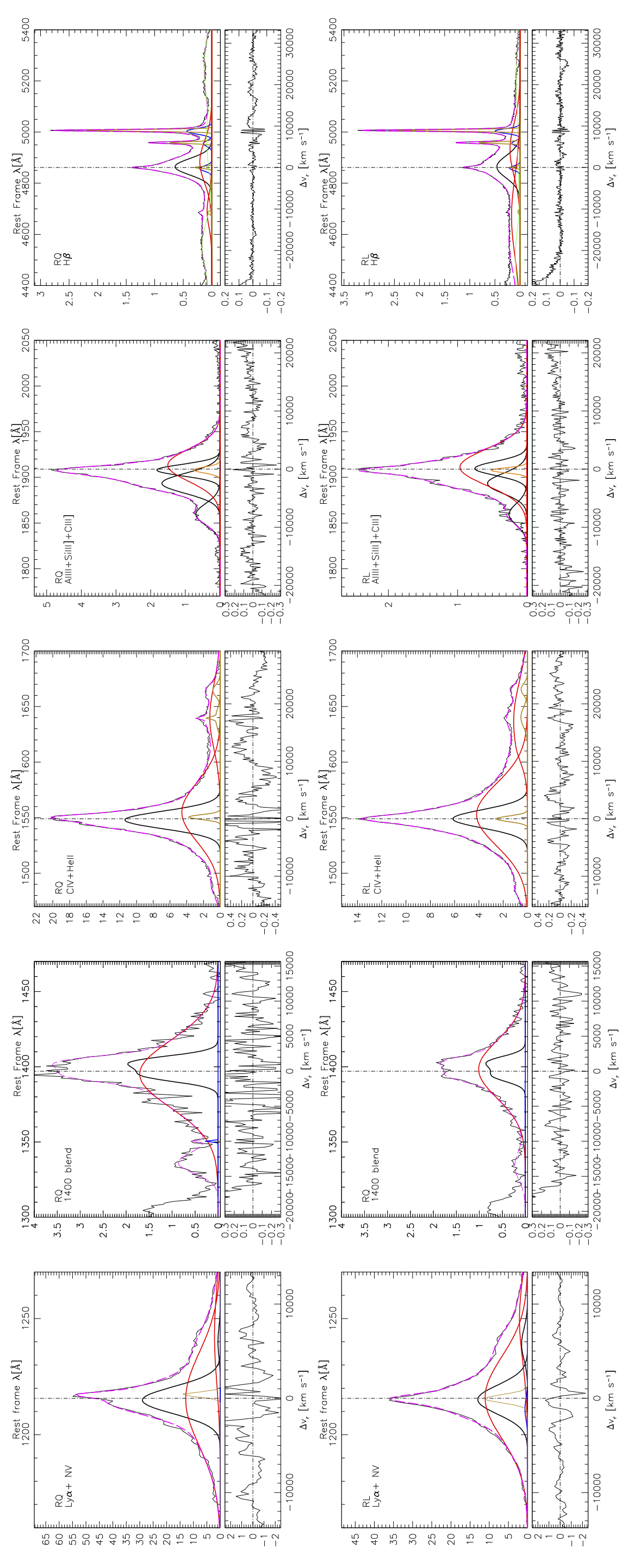} 
  \caption{Analysis of prominent blends helpful for metallicity diagnostics. Continuum subtracted spectra are shown in the rest frame, with the blend ordered with an increasing wavelength from left to right: the \lya\ + \nv\ blend,  \siiv + \oiv;  \civ\ + \heii; the 1900\AA\ blend made up mainly of \aliii, \siiii, \ciii; the \hb\ + \oiiiopt\ region.  Dashed magenta line: model spectrum; thick solid line: broad components; red thick line: very broad components;  thin orange lines: narrow components of \hb\ and \oiii; blue lines: blue-shifted components. Green lines trace the scaled and broadened \feii\ emission template. The lower panels show the observed minus model residuals in the radial velocity scale.}
\label{fig:medians} 
\end{figure}

The composite average spectra are shown in Figure \ref{fig:combined}. %The area shaded {in} grey is covered by fewer spectra and is therefore not used for the analysis. 
The spectral similarity between the two classes already evident in Figure \ref{fig:combined} is confirmed by the profile comparison of Fig. \ref{fig:medians}. Both spectra are \feii\ weak; show weak \nv; \aliii\ is weak while the \ciii\ line is by far the most prominent in the 1900 \AA\ blend encompassing \siiii\ and \ciii\ along with \aliii. All lines can be successfully decomposed into a BC and a VBC. Apparently, there is no VBC in the \aliii\ and \siiii\ lines and in \feii\. No prominent BC is observed in the \heii\ optical and UV lines, at variance with \hb, \lya, \civ, and the other lines. The absence of a prominent BC in Helium lines is typical \citep{marzianisulentic93}, albeit the implications for the BLR structure are not obvious even on a qualitative basis. Table \ref{tab:lines} reports the line fluxes normalized to \hb. In view of the heuristic decomposition approach, fluxes are reported for the BC, VBC, and sum of the two components, i.e., for the full broad profile (total flux). Each value in this table has been normalized by the corresponding component of \hb. Error analysis has been performed empirically or on the basis of previous analyses (\hb\ and \feii: Refs. \cite{marzianietal03a,marzianietal22}; \civ: \citep{sulenticetal17}; blend at 1900 \AA: Refs. \cite{garnicaetal22,marzianietal22}; \nv: Ref. \cite{sniegowskaetal21}).  The undetected \heii\ BC was considered in the models as \heii/\hb $\lesssim $ 0.1. A more refined treatment of uncertainties is deferred to future work.  

 \begin{table}
\begin{center}
\caption{Normalised emission line  intensities \label{tab:lines}}\tabcolsep=5pt
  \begin{tabular}{llcc} \hline\hline
  Line  Id. &   Comp.  &  $I/I_{\mathrm{H}\beta}$ & $I/I_{\mathrm{H}\beta}$ \\  
& & RQ & RL  \\ 
\hline
\nv\	&	BC	&	0.257	$\pm$	0.081		&	0.450	$\pm$	0.142	 	\\
	&	VBC	&	2.604	$\pm$	0.701		&	1.646	$\pm$	0.792	 	\\
	&	Total	&	1.323	$\pm$	0.402		&	1.053	$\pm$	0.320	 	\\
 						\hline						
\siiv + \oiv\	&	BC	&	0.667	$\pm$	0.149		&	0.294	$\pm$	0.053	 	\\
	&	VBC	&	2.237	$\pm$	0.403		&	0.961	$\pm$	0.173	 	\\
	&	Total	&	1.380	$\pm$	0.218		&	0.630	$\pm$	0.100	 	\\
 								\hline					
\civ\	&	BC	&	4.382	$\pm$	0.620		&	2.545	$\pm$	0.360	 	\\
	&	VBC	&	7.158	$\pm$	1.012		&	5.840	$\pm$	0.826	 	\\
	&	Total	&	5.643	$\pm$	0.399		&	4.207	$\pm$	0.297 \\
 					  \hline								
\heiiuv\	&	BC	&	0.228	$\pm$	0.072		&	0.238	$\pm$	0.075	 	\\
	&	VBC	&	2.099	$\pm$	0.469		&	1.817	$\pm$	0.406	 	\\
	&	Total	&	1.078	$\pm$	0.222		&	1.034	$\pm$	0.213	 	\\
						\hline							
\aliii\	&	BC	&	0.388	$\pm$	0.123		&	0.150	$\pm$	0.046	 	\\
	&	Total	&	0.485	$\pm$	0.100		&	0.156	$\pm$	0.063	\\
				\hline									
\ciii\	&	BC	&	0.556	$\pm$	0.124		&	0.291	$\pm$	0.064	\\
	&	VBC	&	1.548	$\pm$	0.346		&	0.951	$\pm$	0.213	\\
	&	Total	&	1.007	$\pm$	0.113		&	0.624	$\pm$	0.070	\\
						\hline							
\feiiq\	&	Total	&	0.344	$\pm$	0.088		&	0.175	$\pm$	0.088	\\
	 		 			\hline		 				 	
\heii\	&	BC	&	$\lesssim 0.1$	 					&	$\lesssim 0.1$	 				\\
	&	VBC	&	0.299	$\pm$	0.152		&	$\lesssim$		1.299	\\
	&	Total	&	0.136	$\pm$	0.068		&	$\lesssim$		0.655	\\
	\hline
  \end{tabular}
  \end{center}
\footnotesize{ }  
\end{table} 

Spectral differences between the two classes are not striking on a qualitative basis. We note a stronger redward asymmetry in RL sources. This is, however, a result known for a long time \citep{marzianietal03b,punsly10}. The optical \feii\ emission is twice as strong in RQ than in RL; this is also consistent with previous results \cite{marzianietal21a}. The intensity ratios most commonly used in $Z$\ estimates are marginally higher for RQ than for RL. \siiv+\oiv/\civ, $\approx$ 0.24 vs 0.15 for RQ and RL, respectively. Similarly, \aliii/\ciii\ $\approx$ 0.2 vs 0.12; \nv/\hb $\approx$ 1.32 vs 1.05, if total fluxes are considered.  

\begin{figure}[t!]
%\centering
\includegraphics[width=6.425 cm]{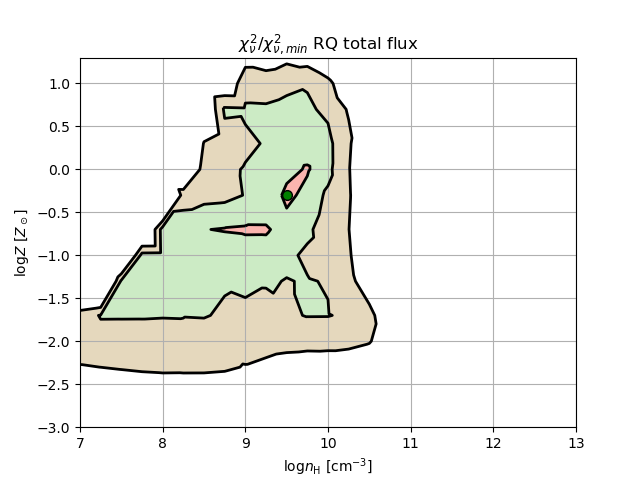}
\hspace{-0.6cm}
\includegraphics[width=6.425 cm]{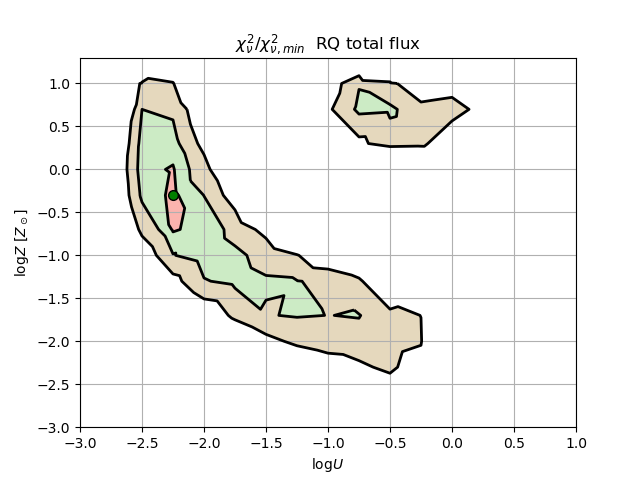}\\
\includegraphics[width=6.425 cm]{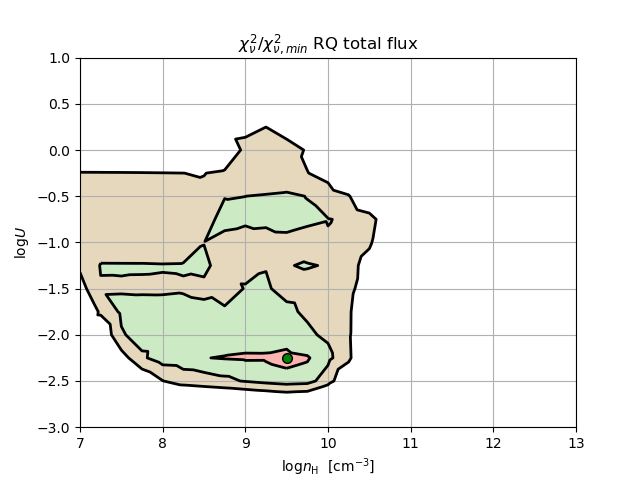}\\
\includegraphics[width=6.425 cm]{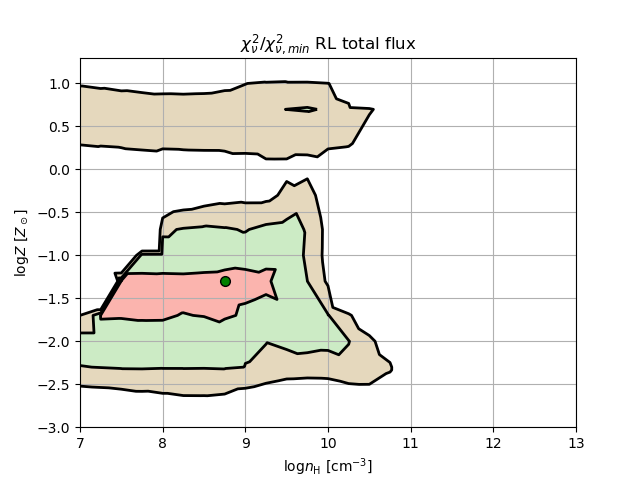}
\hspace{-0.6cm}
\includegraphics[width=6.425 cm]{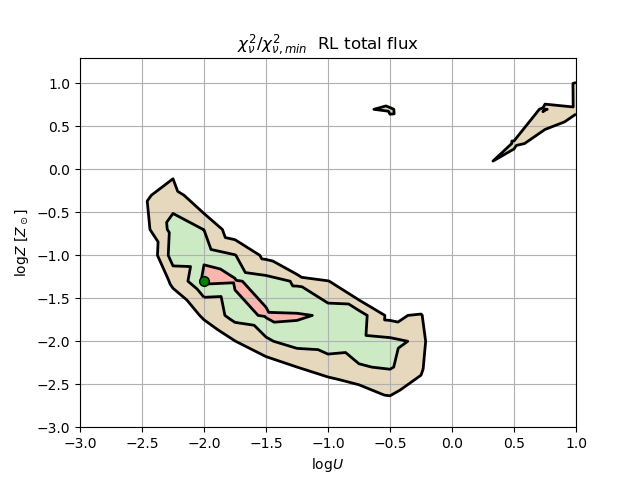}\\
\includegraphics[width=6.425 cm]{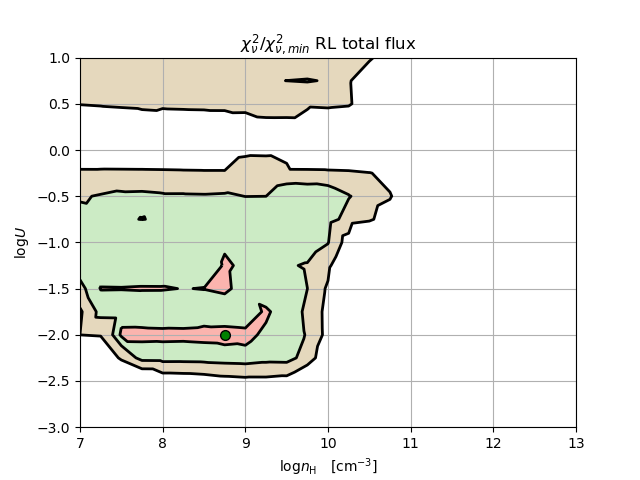}\\
 \caption{Projections of the 3D parameter space \nh, $U$, $Z$. Isophotal contour traces the 1,2,3$\sigma$\ confidence levels; the green spot identifies the values of \nh, $U$, $Z$ yielding the minimum $\chi_\nu^2$. The upper panels are for RQ, lower panels are for RL.  Regions of the parameter space where \oiii/\hb $>0.1$\ have been assigned a very high $\chi_\nu^2$.\label{fig:chi2}  }
\end{figure}     

We computed the { normalized} $\chi^2_\nu$\ in the following form, to constrain the value of the metallicity from the values of the diagnostic ratios. For each spectrum $k$, and for each component $c$, we can write (following Ref. \cite{sniegowskaetal21}):
\begin{eqnarray}
\chi^2_{\nu,\mathrm{kc}}(n_\mathrm{H},U,Z) & = & \frac{1}{n_\mathrm{f}}\sum_\mathrm{{i}}   \left(\frac{R_\mathrm{kci}- R_\mathrm{kci,mod}(n_\mathrm{H},U,Z)}{\delta R_\mathrm{kci}}\right)^2    \label{eq:chi}
\end{eqnarray}
where the summation is made over the available diagnostic ratios. The number of degrees of freedom $n_\mathrm{f}$\ is 8 for the composites and 7 for the case of PKS 0226-038 presented in \S\ \ref{pks0226}.   The  $\chi^2_\nu$ is computed to the results of the {{\em Cloudy}} simulations as a function of $U$, \nh, and $Z$\ (subscript `mod'). { The following { independent} intensity ratios were} considered in the summation:  \nv/\civ, \civ/\heiiuv, \civ/\hb, \siiv+\oiv/\civ,   \aliii/\civ, \aliii/\siiii, \aliii/\ciii, \heii/\hb, \feiiq/\hb. The ratios were computed for BC, VBC, and total line intensity. Non-detections were treated as upper limits.

 Fig. \ref{fig:chi2} shows the projections of the  3D space  $U$, \nh, $Z$. Each contour in the plane braces elements of the grid of  {{\em Cloudy}} parameter space that is consistent with the minimum $\chi^2_\nu$\ within the uncertainties at  {1$\sigma$, 2$\sigma$, and 3$\sigma$ confidence levels}.  

The distribution of the data points is constrained in a  range of $U$, \nh, $Z$, at very low density, relatively high ionization, and low metallicity. Within the limit in  $U$, {and} \nh, the distribution of $Z$\ is slightly subsolar for the RQ ($Z \approx 0.5 Z_\odot$),  and even more subsolar for the radio loud, ($Z \approx 0.1 Z_\odot$). There is a relatively broad range of density and ionization parameters and metallicities that are consistent within $2 \sigma$ confidence level from the minimum $\chi^2$. This is probably a reflection of the stratified nature of the emitting region in Population B and the intrinsic heterogeneity of the composites.  

The values reported in Table \ref{tab:results} for the sub-regions confirm the validity of the virial scenario, with $U$\ increasing by a factor $\gtrsim 10$ between BLR and VBLR. The derived density is not changing strongly, as implied by the strong \ciii\ VBC. { Since the distance from the center of gravity scales with the inverse of the velocity dispersion squared (i.e., $r \propto$ FWHM$^{-2}$, we might expect that $U_\mathrm{}/U_\mathrm{BLR} \sim $ FWHM$_\mathrm{VBLR}^4$/FWHM$_\mathrm{BLR}^4 \sim (2.32 - 2.55)^4 \sim 29 - 42$. The $U$\ values reported in Table \ref{tab:results} indicate ratios between 1 -- 2 orders of magnitudes, consistent with the ones expected from a virial velocity field.}

The $Z$\ values all fall in ranges that are consistent. The results for the sub-regions should be seen with some care, as the VBC measurements are difficult for \heii, \heiiuv, \nv. %For instance, the ratio \nv/\heii\ BC is the ratio between two weak components affected by noise and blended with a much stronger signal from the \lya\ red wing and was also not considered.  
%In addition, no constraints on the \feii\ VBC are included. 
The fairly high metallicity value for RQ VBLR is due to several line ratios being consistently higher than for the RL case: \civ/\hb,  (\siiv+\oiv)/\civ, (\siiv+\-\oiv)/\-\heiiuv, and \nv/\heiiuv\ are between 30\%\ and a factor 2 higher for the VBLR of RQ than of RL. This epitomizes the need for very accurate measurements on the line profiles. 

{These results can be compared with the ones derived from the analysis of the Sloan composite spectrum \citep{vandenberketal04}. Values of the intensity ratios reported in their Table 2, }
%(\nv/\heiiuv $\approx$ 4.72,	\nv/\civ $\approx$ 0.1, \civ/\hb $\approx$  2.92, \siiv+\oiv)/\civ $\approx$ 0.35,	\civ/\heiiuv $\approx$ 48.54,	\aliii/\civ $\approx$ 0.01, \aliii/\siiii $\approx$ 2.11, \aliii/\ciii $\approx$ 0.02, \heii/\hb\ $\approx$ 0.02,\feiiq/\hb $\approx$ 0.43),} 
{ and the assumption of typical  10\%\ uncertainties for most ratios, do not provide any constraint on $Z$, and loose constraints on density ($\log$\nh $\sim 9.5^{+1.5}_{-2.5}.$)  and ionization ($\log U \sim -1.75 - -0.5$). Averaging the full dataset of a large color-based spectroscopic survey presents major hindrances.  %The median composites weight in different ways  $Z$-sensitive line ratios depending on their dispersion along the sequence,
Spectral bins are likely to be associated with intrinsic differences in $Z$. Building a composite combines all survey spectra with a weight proportional to the relative prevalence of each spectral type \citep[e.g., ][]{sulenticmarziani15}. Since the most populated spectral bins along the sequence  is  B1, with A2 the second most populated \citep{marzianietal13a,shenho14}, the appearance of the composite spectrum is resembling the Pop. B RQ composite is some key features, and A2 in others. For example, the moderate \feii/\hb\ and \siiv+\oiv)/\civ\ values are consistent with solar or supersolar supersolar $Z$.   
The composite is yielding line ratios that reflect a combination of physical properties from  individual spectra; on the other hand, it does not reflect the spectral diversity of a large population of type-1 AGN, as  extreme sources are just a minority ($\lesssim 5$\%; Fig. \ref{fig:ms}).  }

  \begin{table}
\begin{center}
\caption{ Derived values of $U$, $Z$, \nh\ and $1\sigma$\ ranges$^a$ \label{tab:results}}
\tabcolsep=5pt
  \begin{tabular}{cc||c|c||c|c||c|c||} \hline\hline 
 \multicolumn{1}{c}{Class} & \multicolumn{1}{c}{Region} &  \multicolumn{1}{c}{$\log U$} &  \multicolumn{1}{c}{$\Delta\log U $}  & \multicolumn{1}{c}{$\log Z$} & \multicolumn{1}{c}{$\Delta\log Z $}  &\multicolumn{1}{c}{$\log$\nh} &  \multicolumn{1}{c}{$\Delta\log$ \nh} \\  
 \hline
RQ & Tot. &   -2.25 &-2.25 -- -2.25  & -0.30& -0.70 --  0.00   &  9.50 & 8.50  -- 9.75   \\
RQ & BLR  &    -2.25 &-2.25 -- -1.75  & 0.30 &-0.70 --  1.00 &  10.25& 9.25 -- 10.75 \\
RQ & VBLR &   0.00   & 0.00 --  0.00  &  0.70 & 0.70 --  0.70 &  9.50 & 9.50  -- 9.75  \\
RL & Tot. &  -2.00 & -2.00 -- -0.75 &   -1.30 &  -2.00 --  -1.30 &  8.75  & 7.00 --  9.75  \\

RL & BLR  &  -1.50 & -2.00 -- -0.75 &   -1.70 &  -2.00 --  -1.00 & 10.25  & 8.75 -- 10.50  \\
RL & VBLR &  -0.75 & -1.25 -- -0.25 &   -2.00 &  -2.00 --  -1.70 &  7.75  & 7.00 -- 10.25  \\
	\hline
  \end{tabular}
  \end{center}
\footnotesize{ $^a$: ionization parameter $U$, abundance $Z$\ in solar units, and Hydrogen particle density \nh\ in units of cm$^{-3}$. The ranges are defined by the limiting elements  of the model grid that are compatible with the minimum $\chi^2_\nu$\ within 1 $\sigma$\ confidence level.}  
\end{table} 

\subsection{A typical RL source at high $z$}
\label{pks0226}
\begin{figure}[t!]
\includegraphics[scale=1.1]{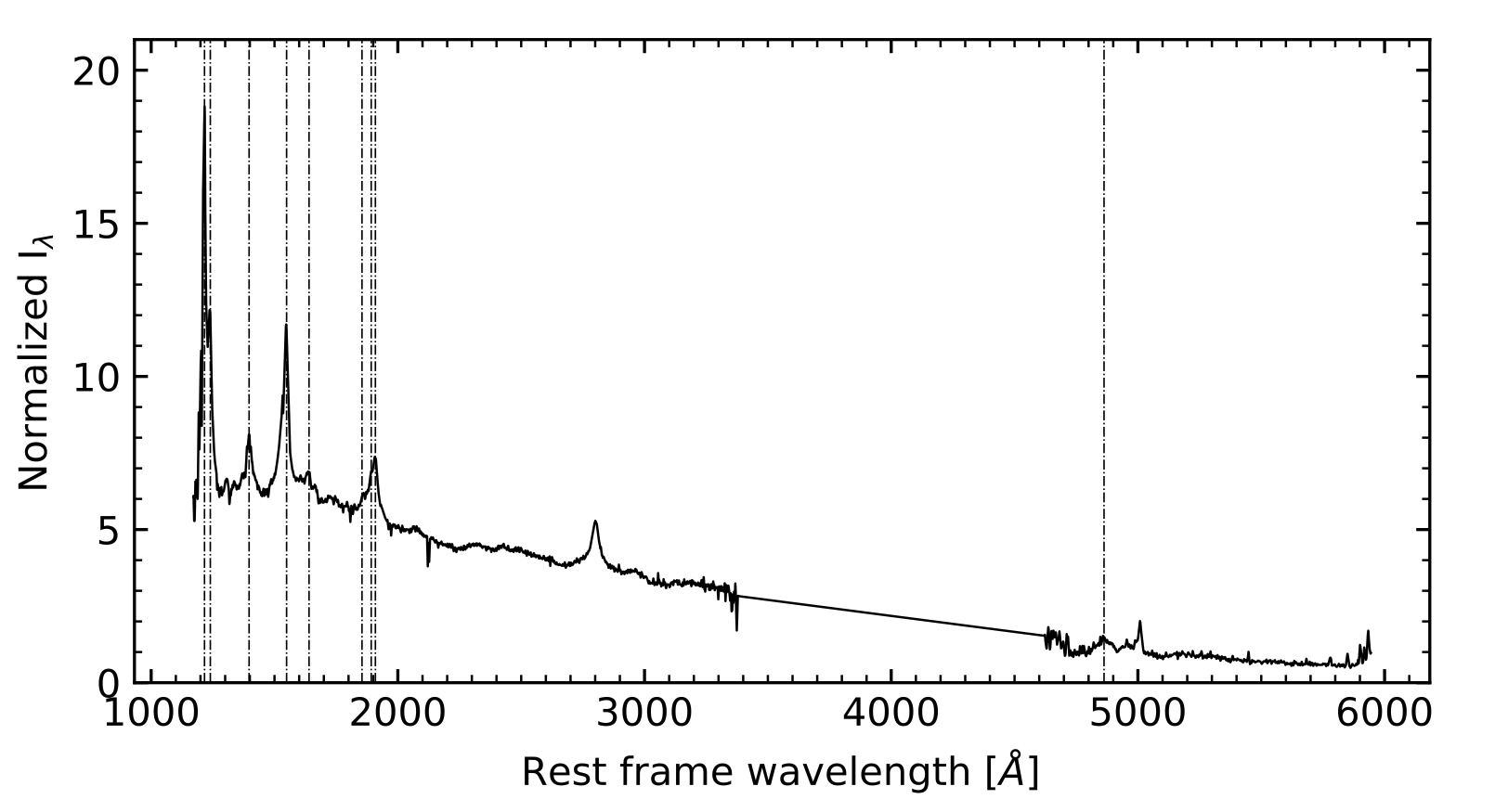} 
 \caption{The spectrum of PKS 0226-038 in the optical and the UV domain. The dot-dashed lines identify the main emission features included in the non-linear $\chi^2$\ fitting procedures. The straight line traces a range where data are unavailable for this source.}
\label{fig:pks0226} 
\end{figure}

PKS 0226--038 is a luminous jetted source at the cosmic noon ($z \approx$ 2.06922; $M_\mathrm{B} \approx -28.0$, and Kellermann's radio loudness parameter \cite{kellermannetal89}, $R_\mathrm K \sim 10^3$). Its optical and UV spectrum
%\footnote{Rest frame UV and optical data will be presented in a forthcoming survey on the spectral properties of RL quasars at high $z$.   } 
is markedly different from the { Population B} composite at low $z$: \ciii\ is weaker, \feii\ stronger, and \aliii\ is almost as strong as \civ\ (Fig. \ref{fig:pks0226}). No diagnostic based on \nv\ and \heii\ is available in this case. The measured diagnostic ratios are reported in Table \ref{tab:pks0226}. The uncertainties were assumed to be $\approx 10$\%\ for the intensities normalized to \hb\ save for \civ/\hb\ that is affected by a re-scaling ($\approx$ 20\%) and for \feiiq\ for which $\delta$\rfe = 0.1 was assumed. {In this case, the \feiiq\ (Fig. \ref{fig:pks0226}) is partly missing because of a telluric absorption. The \feii\ optical emission is well-represented by a "solid" template with fixed multiplet ratios.} 

\begin{table}
\begin{center}
\caption{Diagnostic line ratios measured on the PKS 0226-038 spectrum$^a$ \label{tab:pks0226}}\tabcolsep=5pt
  \begin{tabular}{lc} \hline\hline
  Ratio  Id. &    Value     \\    \hline
\civ/\heiiuv   & 4.436 $\pm$	0.627 \\
(\siiv + \oiv)/\civ\	 	&	0.361		$\pm$0.051\\	
(\siiv + \oiv)/\heiiuv	 	&	 1.603	 $\pm$	 0.227\\
\civ/\hb	 	&	3.460$\pm$	0.720		\\
\aliii/\civ\ & 0.986  $\pm$	0.139\\
\aliii/\siiii\ & 0.642$\pm$	0.091\\
\siiii/\ciii\  & 6.034$\pm$	0.853\\
\rfe\	 	&	0.62	$\pm$	0.10	\\
\hline\\
  \end{tabular}
  \end{center}
\footnotesize{$^a$Line ratios refer  to the total flux of the full broad line profiles. } 
\end{table} 
The derived  values of the metallicity are well constrained with minimum $\chi^2$ \ obtained for $\log Z/Z_\odot \approx 1$, with $1\sigma$ uncertainty {equal to} 0.3 dex. The ionization parameter $\log U \approx -2.75$\ and density $\log $\nh\  $\sim 11.75$ [cm$^{-3}$] are significantly lower and higher {with respect to} the values at low-$z$. The three parameters reflect the moderate \rfe, high \aliii/\civ\ (yielding low ionization and high $Z$), and low \ciii/\siiii\ (implying  a high density).

 \section{Discussion: metal enrichment along the quasar Main Sequence}
\label{discussion}

 Several basic results emerge from the present analysis: 

\begin{description}
     \item[1] RQ Pop. B sources show slightly sub-solar or solar metallicities;    \item in the same spectral bin along the MS, RL sources show definitely sub-solar   chemical abundances, lower with respect to RQ. The difference is not marginal  and is supported by several diagnostic indicators consistently observed to be lower in RL sources. { It is also consistent with the location of the RL template in the MS, displaced toward broader \hb\ and lower \rfe\ with respect to the RQ one. } 
 \item[2] there is an important difference in properties at the extremes of the E1 MS: here, we focused only on the low-\rfe\ extreme deriving sub-solar metallicity. However, highly-supersolar $Z$\ estimates were derived at the other end of the MS, with \rfe$\gtrsim 1$ \ \citep{sniegowskaetal21,garnicaetal22}. 
  \end{description}

The first result is not completely unexpected, since a similar analysis yielded slightly sub-solar abundances for NGC 1275 \citep{punslyetal18a} and Pop. B sources are known to have similar intensity ratios \citep{marzianietal10}.

The origin of the RQ and RL $Z$\ difference is likely associated with a difference in sample/host galaxy evolutionary history. At the modest accretion rates required to sustain the low Eddington ratio of B1 sources, there is no need to invoke "cataclysmic" events such as recent, major wet mergers to explain the AGN luminosity via super-Eddington accretion rates \citep{donofriomarziani18}, and accretion material could be provided via stellar mass loss in early-type galaxies \citep{padovanimatteucci93}. The general impression of low-$z$\ Population B is that of fairly evolved systems, past their prime of accretion events \citep{fraix-burnetetal17}. This impression, is further reinforced in the case of radio-loud sources, where sub-parsec binary black holes at the stable end of their in-spiral phase are relatively frequent \citep{krauseetal19}.  

 The gradient in chemical composition is also no surprise. While it is possible to account for an increase in \feii\ along the sequence only based on ionization degree and density \cite{marzianietal01}, a thorough quantitative assessment requires also a change in chemical composition \cite{pandaetal19}. We pass from sources where singly ionized iron emission is barely detected to sources where it dominates the thermal balance of the emitting regions, overwhelming the entire Balmer line emission.

The case of the high-$z$\ RL source PKS 0226--038 signifies the co-existence of powerful relativistic ejections and BLR physical conditions typical of sources radiating at \lledd $\gtrsim 0.1  - 0.2$\ (Population A according to Ref. \cite{sulenticetal00a}). It is known for a long time that the radio-loud fraction among AGN increases with redshift \cite{padovanietal93,lafrancaetal94}. However, the realization of the high prevalence of unambiguously jetted AGN at $z \gtrsim 1$ radiating at high \lledd\ is a recent result. The Eddington ratio of PKS 0226-038 is $\log$\lledd $\approx -0.43$. It belongs to the spectral type B2, that include sources in the \rfe\ range 0.5 -- 1, and with moderate accretion rate. Strong \feii\ emission is a rather rare occurrence among RLs at low-$z$\ (especially in Fanaroff-Riley II sources, \citep{zamfiretal08}), it is frequently found at high redshift  (Ref. \cite{deconto-machadoetal22}, and article in preparation), along with higher \lledd. This is probably a consequence of the increase in average accretion rate at the epochs corresponding to the cosmic noon \citep{cavalierevittorini00,marconietal04,hopkinsetal06}, and much beyond.  Indeed, highly accreting jetted sources are being discovered up to redshift $\approx 6$\ \citep{banadosetal21,ighinaetal23}.

\section{Conclusions}
     
This pilot work revealed a population of quasars with slightly sub-solar abundances. We can infer a metallicity trend along the quasar MS on the basis of previous results and of the analysis presented in this paper, supporting the assumptions in the study of Panda et al. \cite{pandaetal19}, although the MS-oriented analysis has been yet focused only on the spectral types, B1 and extreme Population A. Full analysis for individual spectral types isolated along the MS \citep{sulenticetal02} is still needed. Special attention should also be devoted to high-$z$\ high-$L$ sources. At high-$z$, sample selection effects could be even more important, excluding a population of sources on the basis of their low accretion rate \citep{sulenticetal14}. At high luminosities, wind effects are expected to be maximized and may introduce another factor in the chemical enrichment of the broad-line emitting gas.

%%%%%%%%%%%%%%%%%%%%%%%%%%%%%%%%%%%%%%%%%%
\authorcontributions{{All authors contributed equally to this paper.}}

%%%%%%%%%%%%%%%%%%%%%%%%%%%%%%%%%%%%%%%%%%
%\funding{{P.M.} wishes to thank Ascensi{o}n del Olmo and Jaime Perea for the allotted time on their departmental server {\tt hypercat} at IAA-CSIC.  }

%%%%%%%%%%%%%%%%%%%%%%%%%%%%%%%%%%%%%%%%%%
\acknowledgments{SP acknowledges the financial support from the Conselho Nacional de Desenvolvimento Científico e Tecnológico (CNPq) Fellowship (164753/2020-6) and the computational facility at Nicolaus Copernicus Astronomical Center in Warsaw, Poland where the {\em Cloudy} simulations presented in this work were performed.}

%%%%%%%%%%%%%%%%%%%%%%%%%%%%%%%%%%%%%%%%%%
\conflictsofinterest{The authors declare no conflict of interest.} 

%%%%%%%%%%%%%%%%%%%%%%%%%%%%%%%%%%%%%%%%%%
%% optional
 
\abbreviations{The following abbreviations are used in this manuscript:\\
\noindent 
\begin{tabular}{@{}ll}
AGN& Active Galactic Nucleus\\
BC & Broad Component \\
BLR& Broad Line Region\\
%{CD}& {Core dominated} \\
%{FR-I}& {Fanaroff-Riley I} \\ 
%{FR-II}& {Fanaroff-Riley II} \\
%FIZ & Fully Ionized Zone\\
%DESI& Dark Energy Spectroscopic Instrument \\
%ESC& Eddington Standard Candles\\
FWHM & Full Width Half-Maximum\\
%HIL& High-Ionization Line\\
%LIL& Low-Ionization Line\\
%MDPI& Multidisciplinary Digital Publishing Institute\\
MS& Main Sequence\\
NLSy1 & Narrow-Line Seyfert 1\\
%PIZ & Partially Ionized Zone \\
{RL} & {Radio loud} \\
{RQ} & {Radio quiet} \\
{SDSS} & {Sloan Digital Sky Survey}\\
{SED} & {Spectral energy distribution}\\
{S/N} & {Signal-to-noise ratio}\\
VBC & Very Broad Component\\
VBLR & Very Broad Line Region \\
%VBE& Virial Broadening Estimator\\
\end{tabular} }

\vfill\eject\pagebreak\newpage

%=====================================
% References, variant B: external bibliography
%=====================================

\externalbibliography{yes}
%\begin{thebibliography}{999}
%\providecommand{\natexlab}[1]{#1}

%\end{thebibliography}
%
%%%%%%%%%%%%%%%%%%%%%%%%%%%%%%%%%%%%%%%%%%
%% optional
%\sampleavailability{Samples of the magic compounds needed to turn quasars into standard candles are {\em not} available from the authors.}

%% for journal Sci
%\reviewreports{\\
%Reviewer 1 comments and authors’ response\\
%Reviewer 2 comments and authors’ response\\
%Reviewer 3 comments and authors’ response
%}
 %\setcounter{figure}{0}
%The Appendix shows the results of the {\tt CLOUDY} simulations in the case of the restricted ranges 4500 -- 4590 \AA\ and 5200 -- 5300 \AA\ ($\tilde{B}$\ and $\tilde{R}$\ through the paper). 

%\section{Restricted range}
%\label{app:restr}

%%%%%%%%%%%%%%%%%%%%%%%%%%%%%%%%%%%%%%%%%%

\begin{thebibliography}{999}

\bibitem[{Netzer}(1990)]{netzer90}
{Netzer}, H.
\newblock {AGN emission lines.}
\newblock  Active Galactic Nuclei; {R.~D.~Blandford, H.~Netzer, L.~Woltjer,
  T.~J.-L.~Courvoisier, \& M.~Mayor}., Ed.,  1990, pp. 57--160.

\bibitem[{Peterson}(1997)]{peterson97}
{Peterson}, B.M.
\newblock {\em {An Introduction to Active Galactic Nuclei}}; Cambridge
  University Press,  1997.

\bibitem[{Osterbrock} and {Ferland}(2006)]{osterbrockferland06}
{Osterbrock}, D.E.; {Ferland}, G.J.
\newblock {\em Astrophysics of gaseous nebulae and active galactic nuclei};
  University Science Books: Mill Valley, CA,  2006.

\bibitem[{Kaler}(1997)]{kaler97}
{Kaler}, J.B.
\newblock {\em {Stars and their Spectra, An Introduction to the Spectral
  Sequence}}; Cambridge University Press,  1997.

\bibitem[{Peterson}(2014)]{peterson14}
{Peterson}, B.M.
\newblock {Measuring the Masses of Supermassive Black Holes}.
\newblock {\em SpScieRev} {\bf 2014}, {\em 183},~253--275.
\newblock
  doi:{\changeurlcolor{black}\href{https://doi.org/10.1007/s11214-013-9987-4}{\detokenize{10.1007/s11214-013-9987-4}}}.

\bibitem[{Du} \em{et~al.}(2016){Du}, {Wang}, {Hu}, {Ho}, {Li}, and
  {Bai}]{duetal16a}
{Du}, P.; {Wang}, J.M.; {Hu}, C.; {Ho}, L.C.; {Li}, Y.R.; {Bai}, J.M.
\newblock {The Fundamental Plane of the Broad-line Region in Active Galactic
  Nuclei}.
\newblock {\em \apjl} {\bf 2016}, {\em 818},~L14,
  \href{http://xxx.lanl.gov/abs/1601.01391}{{\normalfont [1601.01391]}}.
\newblock
  doi:{\changeurlcolor{black}\href{https://doi.org/10.3847/2041-8205/818/1/L14}{\detokenize{10.3847/2041-8205/818/1/L14}}}.

\bibitem[{Marziani} \em{et~al.}(2016){Marziani}, {Sulentic}, {Stirpe},
  {Dultzin}, {Del Olmo}, and {Mart{\'{\i}}nez-Carballo}]{marzianietal16}
{Marziani}, P.; {Sulentic}, J.W.; {Stirpe}, G.M.; {Dultzin}, D.; {Del Olmo},
  A.; {Mart{\'{\i}}nez-Carballo}, M.A.
\newblock {Blue outliers among intermediate redshift quasars}.
\newblock {\em \apss} {\bf 2016}, {\em 361},~3,
  \href{http://xxx.lanl.gov/abs/1511.07138}{{\normalfont [1511.07138]}}.
\newblock
  doi:{\changeurlcolor{black}\href{https://doi.org/10.1007/s10509-015-2590-2}{\detokenize{10.1007/s10509-015-2590-2}}}.

\bibitem[Panda \em{et~al.}(2018)Panda, Czerny, Adhikari, Hryniewicz, Wildy,
  Kuraszkiewicz, and {\'{S}}niegowska]{pandaetal18}
Panda, S.; Czerny, B.; Adhikari, T.P.; Hryniewicz, K.; Wildy, C.;
  Kuraszkiewicz, J.; {\'{S}}niegowska, M.
\newblock Modeling of the Quasar Main Sequence in the Optical Plane.
\newblock {\em The Astrophysical Journal} {\bf 2018}, {\em 866},~115.
\newblock
  doi:{\changeurlcolor{black}\href{https://doi.org/10.3847/1538-4357/aae209}{\detokenize{10.3847/1538-4357/aae209}}}.

\bibitem[{Panda} \em{et~al.}(2019){Panda}, {Marziani}, and
  {Czerny}]{pandaetal19}
{Panda}, S.; {Marziani}, P.; {Czerny}, B.
\newblock {The Quasar Main Sequence Explained by the Combination of Eddington
  Ratio, Metallicity, and Orientation}.
\newblock {\em \apj} {\bf 2019}, {\em 882},~79,
  \href{http://xxx.lanl.gov/abs/1905.01729}{{\normalfont
  [arXiv:astro-ph.HE/1905.01729]}}.
\newblock
  doi:{\changeurlcolor{black}\href{https://doi.org/10.3847/1538-4357/ab3292}{\detokenize{10.3847/1538-4357/ab3292}}}.

\bibitem[{Ferland} \em{et~al.}(2020){Ferland}, {Done}, {Jin}, {Landt}, and
  {Ward}]{ferlandetal20}
{Ferland}, G.J.; {Done}, C.; {Jin}, C.; {Landt}, H.; {Ward}, M.J.
\newblock {State-of-the-art AGN SEDs for photoionization models: BLR
  predictions confront the observations}.
\newblock {\em \mnras} {\bf 2020}, {\em 494},~5917--5922,
  \href{http://xxx.lanl.gov/abs/2004.11873}{{\normalfont
  [arXiv:astro-ph.HE/2004.11873]}}.
\newblock
  doi:{\changeurlcolor{black}\href{https://doi.org/10.1093/mnras/staa1207}{\detokenize{10.1093/mnras/staa1207}}}.

\bibitem[{Panda}(2021)]{Panda_PhD-Thesis_2021}
{Panda}, S.
\newblock {Physical Conditions in the Broad-line Regions of Active Galaxies}.
\newblock PhD thesis, Polish Academy of Sciences, Institute of Physics,  2021.

\bibitem[{Boroson} and {Green}(1992)]{borosongreen92}
{Boroson}, T.A.; {Green}, R.F.
\newblock {The Emission-Line Properties of Low-Redshift Quasi-stellar Objects}.
\newblock {\em \apjs} {\bf 1992}, {\em 80},~109.
\newblock
  doi:{\changeurlcolor{black}\href{https://doi.org/10.1086/191661}{\detokenize{10.1086/191661}}}.

\bibitem[{Sulentic} and {Marziani}(2015)]{sulenticmarziani15}
{Sulentic}, J.; {Marziani}, P.
\newblock {Quasars in the 4D Eigenvector 1 Context: a stroll down memory lane}.
\newblock {\em Frontiers in Astronomy and Space Sciences} {\bf 2015}, {\em
  2},~6,  \href{http://xxx.lanl.gov/abs/1506.01276}{{\normalfont
  [1506.01276]}}.
\newblock
  doi:{\changeurlcolor{black}\href{https://doi.org/10.3389/fspas.2015.00006}{\detokenize{10.3389/fspas.2015.00006}}}.

\bibitem[{Sulentic} \em{et~al.}(2000){Sulentic}, {Marziani}, and
  {Dultzin-Hacyan}]{sulenticetal00a}
{Sulentic}, J.W.; {Marziani}, P.; {Dultzin-Hacyan}, D.
\newblock {Phenomenology of Broad Emission Lines in Active Galactic Nuclei}.
\newblock {\em ARA\&A} {\bf 2000}, {\em 38},~521--571.
\newblock
  doi:{\changeurlcolor{black}\href{https://doi.org/10.1146/annurev.astro.38.1.521}{\detokenize{10.1146/annurev.astro.38.1.521}}}.

\bibitem[{Shen} and {Ho}(2014)]{shenho14}
{Shen}, Y.; {Ho}, L.C.
\newblock {The diversity of quasars unified by accretion and orientation}.
\newblock {\em \nat} {\bf 2014}, {\em 513},~210--213,
  \href{http://xxx.lanl.gov/abs/1409.2887}{{\normalfont [1409.2887]}}.
\newblock
  doi:{\changeurlcolor{black}\href{https://doi.org/10.1038/nature13712}{\detokenize{10.1038/nature13712}}}.

\bibitem[{Sulentic} \em{et~al.}(2002){Sulentic}, {Marziani}, {Zamanov},
  {Bachev}, {Calvani}, and {Dultzin-Hacyan}]{sulenticetal02}
{Sulentic}, J.W.; {Marziani}, P.; {Zamanov}, R.; {Bachev}, R.; {Calvani}, M.;
  {Dultzin-Hacyan}, D.
\newblock {Average Quasar Spectra in the Context of Eigenvector 1}.
\newblock {\em ApJL} {\bf 2002}, {\em 566},~L71--L75,
  \href{http://xxx.lanl.gov/abs/arXiv:astro-ph/0201362}{{\normalfont
  [arXiv:astro-ph/0201362]}}.
\newblock
  doi:{\changeurlcolor{black}\href{https://doi.org/10.1086/339594}{\detokenize{10.1086/339594}}}.

\bibitem[{Zamfir} \em{et~al.}(2008){Zamfir}, {Sulentic}, and
  {Marziani}]{zamfiretal08}
{Zamfir}, S.; {Sulentic}, J.W.; {Marziani}, P.
\newblock {New insights on the QSO radio-loud/radio-quiet dichotomy: SDSS
  spectra in the context of the 4D eigenvector1 parameter space}.
\newblock {\em MNRAS} {\bf 2008}, {\em 387},~856--870,
  \href{http://xxx.lanl.gov/abs/0804.0788}{{\normalfont [0804.0788]}}.
\newblock
  doi:{\changeurlcolor{black}\href{https://doi.org/10.1111/j.1365-2966.2008.13290.x}{\detokenize{10.1111/j.1365-2966.2008.13290.x}}}.

\bibitem[{Marziani} \em{et~al.}(2021){Marziani}, {Berton}, {Panda}, and
  {Bon}]{marzianietal21}
{Marziani}, P.; {Berton}, M.; {Panda}, S.; {Bon}, E.
\newblock {Optical Singly-Ionized Iron Emission in Radio-Quiet and
  Relativistically Jetted Active Galactic Nuclei}.
\newblock {\em Universe} {\bf 2021}, {\em 7},~484,
  \href{http://xxx.lanl.gov/abs/2112.02632}{{\normalfont
  [arXiv:astro-ph.GA/2112.02632]}}.
\newblock
  doi:{\changeurlcolor{black}\href{https://doi.org/10.3390/universe7120484}{\detokenize{10.3390/universe7120484}}}.

\bibitem[{Marziani} \em{et~al.}(2003){Marziani}, {Sulentic}, {Zamanov},
  {Calvani}, {Dultzin-Hacyan}, {Bachev}, and {Zwitter}]{marzianietal03a}
{Marziani}, P.; {Sulentic}, J.W.; {Zamanov}, R.; {Calvani}, M.;
  {Dultzin-Hacyan}, D.; {Bachev}, R.; {Zwitter}, T.
\newblock {An Optical Spectroscopic Atlas of Low-Redshift Active Galactic
  Nuclei}.
\newblock {\em ApJS} {\bf 2003}, {\em 145},~199--211.
\newblock
  doi:{\changeurlcolor{black}\href{https://doi.org/10.1086/346025}{\detokenize{10.1086/346025}}}.

\bibitem[{Sulentic} \em{et~al.}(2007){Sulentic}, {Bachev}, {Marziani},
  {Negrete}, and {Dultzin}]{sulenticetal07}
{Sulentic}, J.W.; {Bachev}, R.; {Marziani}, P.; {Negrete}, C.A.; {Dultzin}, D.
\newblock {C IV {\ensuremath{\lambda}}1549 as an Eigenvector 1 Parameter for
  Active Galactic Nuclei}.
\newblock {\em \apj} {\bf 2007}, {\em 666},~757--777,
  \href{http://xxx.lanl.gov/abs/0705.1895}{{\normalfont
  [arXiv:astro-ph/0705.1895]}}.
\newblock
  doi:{\changeurlcolor{black}\href{https://doi.org/10.1086/519916}{\detokenize{10.1086/519916}}}.

\bibitem[{Marziani} \em{et~al.}(2009){Marziani}, {Sulentic}, {Stirpe},
  {Zamfir}, and {Calvani}]{marzianietal09}
{Marziani}, P.; {Sulentic}, J.W.; {Stirpe}, G.M.; {Zamfir}, S.; {Calvani}, M.
\newblock {VLT/ISAAC spectra of the H{$\beta$} region in intermediate-redshift
  quasars. III. H{$\beta$} broad-line profile analysis and inferences about BLR
  structure}.
\newblock {\em A\&Ap} {\bf 2009}, {\em 495},~83--112,
  \href{http://xxx.lanl.gov/abs/0812.0251}{{\normalfont [0812.0251]}}.
\newblock
  doi:{\changeurlcolor{black}\href{https://doi.org/10.1051/0004-6361:200810764}{\detokenize{10.1051/0004-6361:200810764}}}.

\bibitem[{Snedden} and {Gaskell}(2007)]{sneddengaskell07}
{Snedden}, S.A.; {Gaskell}, C.M.
\newblock {The Case for Optically Thick High-Velocity Broad-Line Region Gas in
  Active Galactic Nuclei}.
\newblock {\em ApJ} {\bf 2007}, {\em 669},~126--134.
\newblock
  doi:{\changeurlcolor{black}\href{https://doi.org/10.1086/521290}{\detokenize{10.1086/521290}}}.

\bibitem[{Peterson} and {Ferland}(1986)]{petersonferland86}
{Peterson}, B.M.; {Ferland}, G.J.
\newblock {An accretion event in the Seyfert galaxy NGC 5548}.
\newblock {\em Nature} {\bf 1986}, {\em 324},~345--347.
\newblock
  doi:{\changeurlcolor{black}\href{https://doi.org/10.1038/324345a0}{\detokenize{10.1038/324345a0}}}.

\bibitem[{Brotherton} \em{et~al.}(1994){Brotherton}, {Wills}, {Francis}, and
  {Steidel}]{brothertonetal94a}
{Brotherton}, M.S.; {Wills}, B.J.; {Francis}, P.J.; {Steidel}, C.C.
\newblock {The intermediate line region of QSOs}.
\newblock {\em \apj} {\bf 1994}, {\em 430},~495--504.
\newblock
  doi:{\changeurlcolor{black}\href{https://doi.org/10.1086/174425}{\detokenize{10.1086/174425}}}.

\bibitem[{Sulentic} \em{et~al.}(2000){Sulentic}, {Zwitter}, {Marziani}, and
  {Dultzin-Hacyan}]{sulenticetal00c}
{Sulentic}, J.W.; {Zwitter}, T.; {Marziani}, P.; {Dultzin-Hacyan}, D.
\newblock {Eigenvector 1: An Optimal Correlation Space for Active Galactic
  Nuclei}.
\newblock {\em ApJL} {\bf 2000}, {\em 536},~L5--L9,
  \href{http://xxx.lanl.gov/abs/arXiv:astro-ph/0005177}{{\normalfont
  [arXiv:astro-ph/0005177]}}.
\newblock
  doi:{\changeurlcolor{black}\href{https://doi.org/10.1086/312717}{\detokenize{10.1086/312717}}}.

\bibitem[{Wang} and {Li}(2011)]{wangli11}
{Wang}, J.; {Li}, Y.
\newblock {Strong Response of the Very Broad H{$\beta$} Emission Line in the
  Luminous Radio-quiet Quasar PG 1416-129}.
\newblock {\em \apjl} {\bf 2011}, {\em 742},~L12,
  \href{http://xxx.lanl.gov/abs/1110.4701}{{\normalfont [1110.4701]}}.
\newblock
  doi:{\changeurlcolor{black}\href{https://doi.org/10.1088/2041-8205/742/1/L12}{\detokenize{10.1088/2041-8205/742/1/L12}}}.

\bibitem[{Zamfir} \em{et~al.}(2010){Zamfir}, {Sulentic}, {Marziani}, and
  {Dultzin}]{zamfiretal10}
{Zamfir}, S.; {Sulentic}, J.W.; {Marziani}, P.; {Dultzin}, D.
\newblock {Detailed characterization of H{$\beta$} emission line profile in
  low-z SDSS quasars}.
\newblock {\em \mnras} {\bf 2010}, {\em 403},~1759,
  \href{http://xxx.lanl.gov/abs/0912.4306}{{\normalfont [0912.4306]}}.
\newblock
  doi:{\changeurlcolor{black}\href{https://doi.org/10.1111/j.1365-2966.2009.16236.x}{\detokenize{10.1111/j.1365-2966.2009.16236.x}}}.

\bibitem[{Kriss}(1994)]{kriss94}
{Kriss}, G.
\newblock {Fitting Models to UV and Optical Spectral Data}.
\newblock {\em Astronomical Data Analysis Software and Systems III, A.S.P.
  Conference Series} {\bf 1994}, {\em 61},~437.

\bibitem[{Negrete} \em{et~al.}(2012){Negrete}, {Dultzin}, {Marziani}, and
  {Sulentic}]{negreteetal12}
{Negrete}, A.; {Dultzin}, D.; {Marziani}, P.; {Sulentic}, J.
\newblock {BLR Physical Conditions in Extreme Population A Quasars: a Method to
  Estimate Central Black Hole Mass at High Redshift}.
\newblock {\em ApJ} {\bf 2012}, {\em 757},~62,
  \href{http://xxx.lanl.gov/abs/1107.3188}{{\normalfont
  [arXiv:astro-ph.CO/1107.3188]}}.

\bibitem[{Negrete} \em{et~al.}(2013){Negrete}, {Dultzin}, {Marziani}, and
  {Sulentic}]{negreteetal13}
{Negrete}, C.A.; {Dultzin}, D.; {Marziani}, P.; {Sulentic}, J.W.
\newblock {Reverberation and Photoionization Estimates of the Broad-line Region
  Radius in Low-z Quasars}.
\newblock {\em \apj} {\bf 2013}, {\em 771},~31,
  \href{http://xxx.lanl.gov/abs/1305.4574}{{\normalfont
  [arXiv:astro-ph.CO/1305.4574]}}.
\newblock
  doi:{\changeurlcolor{black}\href{https://doi.org/10.1088/0004-637X/771/1/31}{\detokenize{10.1088/0004-637X/771/1/31}}}.

\bibitem[{Marziani} \em{et~al.}(2015){Marziani}, {Sulentic}, {Negrete},
  {Dultzin}, {Del Olmo}, {Mart{\'{\i}}nez Carballo}, {Zwitter}, and
  {Bachev}]{marzianietal15}
{Marziani}, P.; {Sulentic}, J.W.; {Negrete}, C.A.; {Dultzin}, D.; {Del Olmo},
  A.; {Mart{\'{\i}}nez Carballo}, M.A.; {Zwitter}, T.; {Bachev}, R.
\newblock {UV spectral diagnostics for low redshift quasars: estimating
  physical conditions and radius of the broad line region}.
\newblock {\em \apss} {\bf 2015}, {\em 356},~339--346,
  \href{http://xxx.lanl.gov/abs/1410.3146}{{\normalfont [1410.3146]}}.
\newblock
  doi:{\changeurlcolor{black}\href{https://doi.org/10.1007/s10509-014-2136-z}{\detokenize{10.1007/s10509-014-2136-z}}}.

\bibitem[{Chieffi} and {Limongi}(2013)]{chieffilimongi13}
{Chieffi}, A.; {Limongi}, M.
\newblock {Pre-supernova Evolution of Rotating Solar Metallicity Stars in the
  Mass Range 13-120 M $_{{\ensuremath{\odot}}}$ and their Explosive Yields}.
\newblock {\em \apj} {\bf 2013}, {\em 764},~21.
\newblock
  doi:{\changeurlcolor{black}\href{https://doi.org/10.1088/0004-637X/764/1/21}{\detokenize{10.1088/0004-637X/764/1/21}}}.

\bibitem[{{\'S}niegowska} \em{et~al.}(2021){{\'S}niegowska}, {Marziani},
  {Czerny}, {Panda}, {Mart{\'\i}nez-Aldama}, {del Olmo}, and
  {D'Onofrio}]{sniegowskaetal21}
{{\'S}niegowska}, M.; {Marziani}, P.; {Czerny}, B.; {Panda}, S.;
  {Mart{\'\i}nez-Aldama}, M.L.; {del Olmo}, A.; {D'Onofrio}, M.
\newblock {High Metal Content of Highly Accreting Quasars}.
\newblock {\em \apj} {\bf 2021}, {\em 910},~115,
  \href{http://xxx.lanl.gov/abs/2009.14177}{{\normalfont
  [arXiv:astro-ph.HE/2009.14177]}}.
\newblock
  doi:{\changeurlcolor{black}\href{https://doi.org/10.3847/1538-4357/abe1c8}{\detokenize{10.3847/1538-4357/abe1c8}}}.

\bibitem[{Garnica} \em{et~al.}(2022){Garnica}, {Negrete}, {Marziani},
  {Dultzin}, {{\'S}niegowska}, and {Panda}]{garnicaetal22}
{Garnica}, K.; {Negrete}, C.A.; {Marziani}, P.; {Dultzin}, D.;
  {{\'S}niegowska}, M.; {Panda}, S.
\newblock {High metal content of highly accreting quasars: Analysis of an
  extended sample}.
\newblock {\em \aap} {\bf 2022}, {\em 667},~A105,
  \href{http://xxx.lanl.gov/abs/2208.02387}{{\normalfont
  [arXiv:astro-ph.GA/2208.02387]}}.
\newblock
  doi:{\changeurlcolor{black}\href{https://doi.org/10.1051/0004-6361/202142837}{\detokenize{10.1051/0004-6361/202142837}}}.

\bibitem[{Ferland} \em{et~al.}(2017){Ferland}, {Chatzikos}, {Guzm{\'a}n},
  {Lykins}, {van Hoof}, {Williams}, {Abel}, {Badnell}, {Keenan}, {Porter}, and
  {Stancil}]{ferlandetal17}
{Ferland}, G.J.; {Chatzikos}, M.; {Guzm{\'a}n}, F.; {Lykins}, M.L.; {van Hoof},
  P.A.M.; {Williams}, R.J.R.; {Abel}, N.P.; {Badnell}, N.R.; {Keenan}, F.P.;
  {Porter}, R.L.; {Stancil}, P.C.
\newblock {The 2017 Release Cloudy}.
\newblock {\em \rmxaa} {\bf 2017}, {\em 53},~385--438,
  \href{http://xxx.lanl.gov/abs/1705.10877}{{\normalfont
  [arXiv:astro-ph.GA/1705.10877]}}.

\bibitem[{Laor} \em{et~al.}(1997){Laor}, {Fiore}, {Elvis}, {Wilkes}, and
  {McDowell}]{laoretal97b}
{Laor}, A.; {Fiore}, F.; {Elvis}, M.; {Wilkes}, B.J.; {McDowell}, J.C.
\newblock {The Soft X-Ray Properties of a Complete Sample of Optically Selected
  Quasars. II. Final Results}.
\newblock {\em ApJ} {\bf 1997}, {\em 477},~93--+,
  \href{http://xxx.lanl.gov/abs/arXiv:astro-ph/9609164}{{\normalfont
  [arXiv:astro-ph/9609164]}}.
\newblock
  doi:{\changeurlcolor{black}\href{https://doi.org/10.1086/303696}{\detokenize{10.1086/303696}}}.

\bibitem[{Baldwin} \em{et~al.}(1995){Baldwin}, {Ferland}, {Korista}, and
  {Verner}]{baldwinetal95}
{Baldwin}, J.; {Ferland}, G.; {Korista}, K.; {Verner}, D.
\newblock {Locally Optimally Emitting Clouds and the Origin of Quasar Emission
  Lines}.
\newblock {\em ApJL} {\bf 1995}, {\em 455},~L119+,
  \href{http://xxx.lanl.gov/abs/arXiv:astro-ph/9510080}{{\normalfont
  [arXiv:astro-ph/9510080]}}.
\newblock
  doi:{\changeurlcolor{black}\href{https://doi.org/10.1086/309827}{\detokenize{10.1086/309827}}}.

\bibitem[{Korista} \em{et~al.}(1997){Korista}, {Baldwin}, {Ferland}, and
  {Verner}]{koristaetal97}
{Korista}, K.; {Baldwin}, J.; {Ferland}, G.; {Verner}, D.
\newblock {An Atlas of Computed Equivalent Widths of Quasar Broad Emission
  Lines}.
\newblock {\em ApJS} {\bf 1997}, {\em 108},~401--+,
  \href{http://xxx.lanl.gov/abs/arXiv:astro-ph/9611220}{{\normalfont
  [arXiv:astro-ph/9611220]}}.
\newblock
  doi:{\changeurlcolor{black}\href{https://doi.org/10.1086/312966}{\detokenize{10.1086/312966}}}.

\bibitem[{Guo} \em{et~al.}(2020){Guo}, {Shen}, {He}, {Wang}, {Liu}, {Wang},
  {Sun}, {Yang}, {Kong}, and {Sheng}]{Guoetal2020ApJ}
{Guo}, H.; {Shen}, Y.; {He}, Z.; {Wang}, T.; {Liu}, X.; {Wang}, S.; {Sun}, M.;
  {Yang}, Q.; {Kong}, M.; {Sheng}, Z.
\newblock {Understanding Broad Mg II Variability in Quasars with
  Photoionization: Implications for Reverberation Mapping and Changing-look
  Quasars}.
\newblock {\em \apj} {\bf 2020}, {\em 888},~58,
  \href{http://xxx.lanl.gov/abs/1907.06669}{{\normalfont
  [arXiv:astro-ph.GA/1907.06669]}}.
\newblock
  doi:{\changeurlcolor{black}\href{https://doi.org/10.3847/1538-4357/ab5db0}{\detokenize{10.3847/1538-4357/ab5db0}}}.

\bibitem[{Marziani} and {Sulentic}(1993)]{marzianisulentic93}
{Marziani}, P.; {Sulentic}, J.W.
\newblock {Evidence for a very broad line region in PG 1138+222}.
\newblock {\em ApJ} {\bf 1993}, {\em 409},~612--616,
  \href{http://xxx.lanl.gov/abs/arXiv:astro-ph/9210005}{{\normalfont
  [arXiv:astro-ph/9210005]}}.
\newblock
  doi:{\changeurlcolor{black}\href{https://doi.org/10.1086/172692}{\detokenize{10.1086/172692}}}.

\bibitem[{Marziani} \em{et~al.}(2022){Marziani}, {Olmo}, {Negrete}, {Dultzin},
  {Piconcelli}, {Vietri}, {Mart{\'\i}nez-Aldama}, {D'Onofrio}, {Bon}, {Bon},
  {Deconto Machado}, {Stirpe}, and {Buendia Rios}]{marzianietal22}
{Marziani}, P.; {Olmo}, A.d.; {Negrete}, C.A.; {Dultzin}, D.; {Piconcelli}, E.;
  {Vietri}, G.; {Mart{\'\i}nez-Aldama}, M.L.; {D'Onofrio}, M.; {Bon}, E.;
  {Bon}, N.; {Deconto Machado}, A.; {Stirpe}, G.M.; {Buendia Rios}, T.M.
\newblock {The Intermediate-ionization Lines as Virial Broadening Estimators
  for Population A Quasars}.
\newblock {\em \apjs} {\bf 2022}, {\em 261},~30,
  \href{http://xxx.lanl.gov/abs/2205.07034}{{\normalfont
  [arXiv:astro-ph.GA/2205.07034]}}.
\newblock
  doi:{\changeurlcolor{black}\href{https://doi.org/10.3847/1538-4365/ac6fd6}{\detokenize{10.3847/1538-4365/ac6fd6}}}.

\bibitem[{Sulentic} \em{et~al.}(2017){Sulentic}, {del Olmo}, {Marziani},
  {Mart{\'{\i}}nez-Carballo}, {D'Onofrio}, {Dultzin}, {Perea},
  {Mart{\'{\i}}nez-Aldama}, {Negrete}, {Stirpe}, and {Zamfir}]{sulenticetal17}
{Sulentic}, J.W.; {del Olmo}, A.; {Marziani}, P.; {Mart{\'{\i}}nez-Carballo},
  M.A.; {D'Onofrio}, M.; {Dultzin}, D.; {Perea}, J.; {Mart{\'{\i}}nez-Aldama},
  M.L.; {Negrete}, C.A.; {Stirpe}, G.M.; {Zamfir}, S.
\newblock {What does CIV{$\lambda$}1549 tell us about the physical driver of
  the Eigenvector quasar sequence?}
\newblock {\em \aap} {\bf 2017}, {\em 608},~A122,
  \href{http://xxx.lanl.gov/abs/1708.03187}{{\normalfont [1708.03187]}}.
\newblock
  doi:{\changeurlcolor{black}\href{https://doi.org/10.1051/0004-6361/201630309}{\detokenize{10.1051/0004-6361/201630309}}}.

\bibitem[{Marziani} \em{et~al.}(2003){Marziani}, {Zamanov}, {Sulentic}, and
  {Calvani}]{marzianietal03b}
{Marziani}, P.; {Zamanov}, R.K.; {Sulentic}, J.W.; {Calvani}, M.
\newblock {Searching for the physical drivers of eigenvector 1: influence of
  black hole mass and Eddington ratio}.
\newblock {\em MNRAS} {\bf 2003}, {\em 345},~1133--1144,
  \href{http://xxx.lanl.gov/abs/arXiv:astro-ph/0307367}{{\normalfont
  [arXiv:astro-ph/0307367]}}.
\newblock
  doi:{\changeurlcolor{black}\href{https://doi.org/10.1046/j.1365-2966.2003.07033.x}{\detokenize{10.1046/j.1365-2966.2003.07033.x}}}.

\bibitem[{Punsly}(2010)]{punsly10}
{Punsly}, B.
\newblock {The Redshifted Excess in Quasar C IV Broad Emission Lines}.
\newblock {\em \apj} {\bf 2010}, {\em 713},~232--238,
  \href{http://xxx.lanl.gov/abs/1002.4681}{{\normalfont [1002.4681]}}.
\newblock
  doi:{\changeurlcolor{black}\href{https://doi.org/10.1088/0004-637X/713/1/232}{\detokenize{10.1088/0004-637X/713/1/232}}}.

\bibitem[{Marziani} \em{et~al.}(2021){Marziani}, {Berton}, {Panda}, and
  {Bon}]{marzianietal21a}
{Marziani}, P.; {Berton}, M.; {Panda}, S.; {Bon}, E.
\newblock {Optical Singly-Ionized Iron Emission in Radio-Quiet and
  Relativistically Jetted Active Galactic Nuclei}.
\newblock {\em Universe} {\bf 2021}, {\em 7},~484,
  \href{http://xxx.lanl.gov/abs/2112.02632}{{\normalfont
  [arXiv:astro-ph.GA/2112.02632]}}.
\newblock
  doi:{\changeurlcolor{black}\href{https://doi.org/10.3390/universe7120484}{\detokenize{10.3390/universe7120484}}}.

\bibitem[{Vanden Berk} \em{et~al.}(2004){Vanden Berk}, {Wilhite}, {Kron},
  {Anderson}, {Brunner}, {Hall}, {Ivezi{\'c}}, {Richards}, {Schneider}, {York},
  {Brinkmann}, {Lamb}, {Nichol}, and {Schlegel}]{vandenberketal04}
{Vanden Berk}, D.E.; {Wilhite}, B.C.; {Kron}, R.G.; {Anderson}, S.F.;
  {Brunner}, R.J.; {Hall}, P.B.; {Ivezi{\'c}}, {\v Z}.; {Richards}, G.T.;
  {Schneider}, D.P.; {York}, D.G.; {Brinkmann}, J.V.; {Lamb}, D.Q.; {Nichol},
  R.C.; {Schlegel}, D.J.
\newblock {The Ensemble Photometric Variability of \~{}25,000 Quasars in the
  Sloan Digital Sky Survey}.
\newblock {\em \apj} {\bf 2004}, {\em 601},~692--714,
  \href{http://xxx.lanl.gov/abs/astro-ph/0310336}{{\normalfont
  [astro-ph/0310336]}}.
\newblock
  doi:{\changeurlcolor{black}\href{https://doi.org/10.1086/380563}{\detokenize{10.1086/380563}}}.

\bibitem[{Marziani} \em{et~al.}(2013){Marziani}, {Sulentic}, {Plauchu-Frayn},
  and {del Olmo}]{marzianietal13a}
{Marziani}, P.; {Sulentic}, J.W.; {Plauchu-Frayn}, I.; {del Olmo}, A.
\newblock {Is Mg II 2800 a Reliable Virial Broadening Estimator for Quasars?}
\newblock {\em AAp} {\bf 2013}, {\em 555},~89, 16pp,
  \href{http://xxx.lanl.gov/abs/1305.1096}{{\normalfont
  [arXiv:astro-ph.CO/1305.1096]}}.

\bibitem[{Kellermann} \em{et~al.}(1989){Kellermann}, {Sramek}, {Schmidt},
  {Shaffer}, and {Green}]{kellermannetal89}
{Kellermann}, K.I.; {Sramek}, R.; {Schmidt}, M.; {Shaffer}, D.B.; {Green}, R.
\newblock {VLA Observations of Objects in the Palomar Bright Quasar Survey}.
\newblock {\em \aj} {\bf 1989}, {\em 98},~1195.
\newblock
  doi:{\changeurlcolor{black}\href{https://doi.org/10.1086/115207}{\detokenize{10.1086/115207}}}.

\bibitem[{Punsly} \em{et~al.}(2018){Punsly}, {Marziani}, {Bennert}, {Nagai},
  and {Gurwell}]{punslyetal18a}
{Punsly}, B.; {Marziani}, P.; {Bennert}, V.N.; {Nagai}, H.; {Gurwell}, M.A.
\newblock {Revealing the Broad Line Region of NGC 1275: The Relationship to Jet
  Power}.
\newblock {\em \apj} {\bf 2018}, {\em 869},~143,
  \href{http://xxx.lanl.gov/abs/1810.11716}{{\normalfont [1810.11716]}}.
\newblock
  doi:{\changeurlcolor{black}\href{https://doi.org/10.3847/1538-4357/aaec75}{\detokenize{10.3847/1538-4357/aaec75}}}.

\bibitem[{Marziani} \em{et~al.}(2010){Marziani}, {Sulentic}, {Negrete},
  {Dultzin}, {Zamfir}, and {Bachev}]{marzianietal10}
{Marziani}, P.; {Sulentic}, J.W.; {Negrete}, C.A.; {Dultzin}, D.; {Zamfir}, S.;
  {Bachev}, R.
\newblock {Broad-line region physical conditions along the quasar eigenvector 1
  sequence}.
\newblock {\em \mnras} {\bf 2010}, {\em 409},~1033--1048,
  \href{http://xxx.lanl.gov/abs/1007.3187}{{\normalfont
  [arXiv:astro-ph.CO/1007.3187]}}.
\newblock
  doi:{\changeurlcolor{black}\href{https://doi.org/10.1111/j.1365-2966.2010.17357.x}{\detokenize{10.1111/j.1365-2966.2010.17357.x}}}.

\bibitem[{D'Onofrio} and {Marziani}(2018)]{donofriomarziani18}
{D'Onofrio}, M.; {Marziani}, P.
\newblock {A multimessenger view of galaxies and quasars from now to
  mid-century}.
\newblock {\em Frontiers in Astronomy and Space Sciences} {\bf 2018}, {\em
  5},~31,  \href{http://xxx.lanl.gov/abs/1807.07435}{{\normalfont
  [arXiv:astro-ph.GA/1807.07435]}}.
\newblock
  doi:{\changeurlcolor{black}\href{https://doi.org/10.3389/fspas.2018.00031}{\detokenize{10.3389/fspas.2018.00031}}}.

\bibitem[{Padovani} and {Matteucci}(1993)]{padovanimatteucci93}
{Padovani}, P.; {Matteucci}, F.
\newblock {Stellar Mass Loss in Elliptical Galaxies and the Fueling of Active
  Galactic Nuclei}.
\newblock {\em \apj} {\bf 1993}, {\em 416},~26.
\newblock
  doi:{\changeurlcolor{black}\href{https://doi.org/10.1086/173212}{\detokenize{10.1086/173212}}}.

\bibitem[Fraix-Burnet \em{et~al.}(2017)Fraix-Burnet, Marziani, D'Onofrio, and
  Dultzin]{fraix-burnetetal17}
Fraix-Burnet, D.; Marziani, P.; D'Onofrio, M.; Dultzin, D.
\newblock The Phylogeny of Quasars and the Ontogeny of Their Central Black
  Holes.
\newblock {\em Frontiers in Astronomy and Space Sciences} {\bf 2017}, {\em
  4},~1.
\newblock
  doi:{\changeurlcolor{black}\href{https://doi.org/10.3389/fspas.2017.00001}{\detokenize{10.3389/fspas.2017.00001}}}.

\bibitem[{Krause} \em{et~al.}(2019){Krause}, {Shabala}, {Hardcastle},
  {Bicknell}, {B{\"o}hringer}, {Chon}, {Nawaz}, {Sarzi}, and
  {Wagner}]{krauseetal19}
{Krause}, M.G.H.; {Shabala}, S.S.; {Hardcastle}, M.J.; {Bicknell}, G.V.;
  {B{\"o}hringer}, H.; {Chon}, G.; {Nawaz}, M.A.; {Sarzi}, M.; {Wagner}, A.Y.
\newblock {How frequent are close supermassive binary black holes in powerful
  jet sources?}
\newblock {\em \mnras} {\bf 2019}, {\em 482},~240--261,
  \href{http://xxx.lanl.gov/abs/1809.04050}{{\normalfont
  [arXiv:astro-ph.HE/1809.04050]}}.
\newblock
  doi:{\changeurlcolor{black}\href{https://doi.org/10.1093/mnras/sty2558}{\detokenize{10.1093/mnras/sty2558}}}.

\bibitem[{Marziani} \em{et~al.}(2001){Marziani}, {Sulentic}, {Zwitter},
  {Dultzin-Hacyan}, and {Calvani}]{marzianietal01}
{Marziani}, P.; {Sulentic}, J.W.; {Zwitter}, T.; {Dultzin-Hacyan}, D.;
  {Calvani}, M.
\newblock {Searching for the Physical Drivers of the Eigenvector 1 Correlation
  Space}.
\newblock {\em ApJ} {\bf 2001}, {\em 558},~553--560,
  \href{http://xxx.lanl.gov/abs/arXiv:astro-ph/0105343}{{\normalfont
  [arXiv:astro-ph/0105343]}}.
\newblock
  doi:{\changeurlcolor{black}\href{https://doi.org/10.1086/322286}{\detokenize{10.1086/322286}}}.

\bibitem[{Padovani} \em{et~al.}(1993){Padovani}, {Ghisellini}, {Fabian}, and
  {Celotti}]{padovanietal93}
{Padovani}, P.; {Ghisellini}, G.; {Fabian}, A.C.; {Celotti}, A.
\newblock {Radio-loud AGN and the extragalactic gamma-ray background.}
\newblock {\em \mnras} {\bf 1993}, {\em 260},~L21--L24.
\newblock
  doi:{\changeurlcolor{black}\href{https://doi.org/10.1093/mnras/260.1.L21}{\detokenize{10.1093/mnras/260.1.L21}}}.

\bibitem[{La Franca} \em{et~al.}(1994){La Franca}, {Gregorini}, {Cristiani},
  {de Ruiter}, and {Owen}]{lafrancaetal94}
{La Franca}, F.; {Gregorini}, L.; {Cristiani}, S.; {de Ruiter}, H.; {Owen}, F.
\newblock {Deep VLA Observations of an Optically Selected Sample of
  Intermediate Redshift QSOs and the Optical Luminosity Function of the Radio
  Loud QSOs}.
\newblock {\em \aj} {\bf 1994}, {\em 108},~1548.
\newblock
  doi:{\changeurlcolor{black}\href{https://doi.org/10.1086/117176}{\detokenize{10.1086/117176}}}.

\bibitem[{Deconto-Machado} \em{et~al.}(2022){Deconto-Machado}, {del Olmo},
  {Marziani}, {Perea}, and {Stirpe}]{deconto-machadoetal22}
{Deconto-Machado}, A.; {del Olmo}, A.; {Marziani}, P.; {Perea}, J.; {Stirpe},
  G.M.
\newblock {High-redshift quasars along the Main Sequence}.
\newblock {\em arXiv e-prints} {\bf 2022}, p. arXiv:2211.03853,
  \href{http://xxx.lanl.gov/abs/2211.03853}{{\normalfont
  [arXiv:astro-ph.GA/2211.03853]}}.

\bibitem[{Cavaliere} and {Vittorini}(2000)]{cavalierevittorini00}
{Cavaliere}, A.; {Vittorini}, V.
\newblock {The Fall of the Quasar Population}.
\newblock {\em \apj} {\bf 2000}, {\em 543},~599--610,
  \href{http://xxx.lanl.gov/abs/arXiv:astro-ph/0006194}{{\normalfont
  [arXiv:astro-ph/0006194]}}.
\newblock
  doi:{\changeurlcolor{black}\href{https://doi.org/10.1086/317155}{\detokenize{10.1086/317155}}}.

\bibitem[{Marconi} \em{et~al.}(2004){Marconi}, {Risaliti}, {Gilli}, {Hunt},
  {Maiolino}, and {Salvati}]{marconietal04}
{Marconi}, A.; {Risaliti}, G.; {Gilli}, R.; {Hunt}, L.K.; {Maiolino}, R.;
  {Salvati}, M.
\newblock {Local supermassive black holes, relics of active galactic nuclei and
  the X-ray background}.
\newblock {\em \mnras} {\bf 2004}, {\em 351},~169--185,
  \href{http://xxx.lanl.gov/abs/astro-ph/0311619}{{\normalfont
  [arXiv:astro-ph/astro-ph/0311619]}}.
\newblock
  doi:{\changeurlcolor{black}\href{https://doi.org/10.1111/j.1365-2966.2004.07765.x}{\detokenize{10.1111/j.1365-2966.2004.07765.x}}}.

\bibitem[{Hopkins} \em{et~al.}(2006){Hopkins}, {Hernquist}, {Cox}, {Di Matteo},
  {Robertson}, and {Springel}]{hopkinsetal06}
{Hopkins}, P.F.; {Hernquist}, L.; {Cox}, T.J.; {Di Matteo}, T.; {Robertson},
  B.; {Springel}, V.
\newblock {A Unified, Merger-driven Model of the Origin of Starbursts, Quasars,
  the Cosmic X-Ray Background, Supermassive Black Holes, and Galaxy Spheroids}.
\newblock {\em \apjs} {\bf 2006}, {\em 163},~1--49,
  \href{http://xxx.lanl.gov/abs/astro-ph/0506398}{{\normalfont
  [astro-ph/0506398]}}.
\newblock
  doi:{\changeurlcolor{black}\href{https://doi.org/10.1086/499298}{\detokenize{10.1086/499298}}}.

\bibitem[{Ba{\~n}ados} \em{et~al.}(2021){Ba{\~n}ados}, {Mazzucchelli},
  {Momjian}, {Eilers}, {Wang}, {Schindler}, {Connor}, {Andika}, {Barth},
  {Carilli}, {Davies}, {Decarli}, {Fan}, {Farina}, {Hennawi}, {Pensabene},
  {Stern}, {Venemans}, {Wenzl}, and {Yang}]{banadosetal21}
{Ba{\~n}ados}, E.; {Mazzucchelli}, C.; {Momjian}, E.; {Eilers}, A.C.; {Wang},
  F.; {Schindler}, J.T.; {Connor}, T.; {Andika}, I.T.; {Barth}, A.J.;
  {Carilli}, C.; {Davies}, F.B.; {Decarli}, R.; {Fan}, X.; {Farina}, E.P.;
  {Hennawi}, J.F.; {Pensabene}, A.; {Stern}, D.; {Venemans}, B.P.; {Wenzl}, L.;
  {Yang}, J.
\newblock {The Discovery of a Highly Accreting, Radio-loud Quasar at z = 6.82}.
\newblock {\em \apj} {\bf 2021}, {\em 909},~80,
  \href{http://xxx.lanl.gov/abs/2103.03295}{{\normalfont
  [arXiv:astro-ph.CO/2103.03295]}}.
\newblock
  doi:{\changeurlcolor{black}\href{https://doi.org/10.3847/1538-4357/abe239}{\detokenize{10.3847/1538-4357/abe239}}}.
\bibitem[{Ighina} \em{et~al.}(2023){Ighina}, {Caccianiga}, {Moretti},
  {Belladitta}, {Broderick}, {Drouart}, {Leung}, and {Seymour}]{ighinaetal23}
{Ighina}, L.; {Caccianiga}, A.; {Moretti}, A.; {Belladitta}, S.; {Broderick},
  J.W.; {Drouart}, G.; {Leung}, J.K.; {Seymour}, N.
\newblock {New radio-loud QSOs at the end of the Re-ionization epoch}.
\newblock {\em \mnras} {\bf 2023}, {\em 519},~2060--2068,
  \href{http://xxx.lanl.gov/abs/2212.06168}{{\normalfont
  [arXiv:astro-ph.GA/2212.06168]}}.
\newblock
  doi:{\changeurlcolor{black}\href{https://doi.org/10.1093/mnras/stac3668}{\detokenize{10.1093/mnras/stac3668}}}.
\bibitem[{Sulentic} \em{et~al.}(2014){Sulentic}, {Marziani}, {del Olmo},
  {Dultzin}, {Perea}, and {Alenka Negrete}]{sulenticetal14}
{Sulentic}, J.W.; {Marziani}, P.; {del Olmo}, A.; {Dultzin}, D.; {Perea}, J.;
  {Alenka Negrete}, C.
\newblock {GTC spectra of z {$\approx$} 2.3 quasars: comparison with local
  luminosity analogs}.
\newblock {\em \aap} {\bf 2014}, {\em 570},~A96,
  \href{http://xxx.lanl.gov/abs/1406.5920}{{\normalfont [1406.5920]}}.
\newblock
  doi:{\changeurlcolor{black}\href{https://doi.org/10.1051/0004-6361/201423975}{\detokenize{10.1051/0004-6361/201423975}}}.
\end{thebibliography}
\end{document}